\documentclass[useAMS,usenatbib]{mn2e}
\usepackage{graphicx}

\title[Photosphere-Internal Shock Model of GRBs]{Photosphere-Internal Shock Model of Gamma-Ray Bursts: 
Case Studies of {\it Fermi}/LAT Bursts}
\author[K. Toma, X.-F. Wu, P. M\'{e}sz\'{a}ros]
{K. Toma$^{1,2}$, X.-F. Wu$^{1,2,3}$, and P. M\'{e}sz\'{a}ros$^{1,2,4}$
\thanks{E-mail toma@astro.psu.edu (KT); xfwu@pmo.ac.cn (XFW); nnp@astro.psu.edu (PM)}\\
$^{1}$Department of Astronomy and Astrophysics, Pennsylvania State University, 525 Davey Lab,
University Park, PA 16802, USA\\
$^{2}$Center for Particle Astrophysics, Pennsylvania State University\\
$^{3}$Purple Mountain Observatory, Chinese Academy of Sciences, Nanjing 210008, China\\
$^{4}$Department of Physics, Pennsylvania State University\\
}
\begin{document}


\pagerange{\pageref{firstpage}--\pageref{lastpage}} \pubyear{2010}

\maketitle

\label{firstpage}

\begin{abstract}
Radially inhomogeneous gamma-ray burst (GRB) jets release variable photospheric emission and 
can have internal shocks occurring above the photosphere. We generically formulate a photospheric 
emission model of GRBs including Compton up-scattered photospheric (UP) emission off the electrons
(and positrons) in the internal shocks, and find that the photospheric emission may correspond 
to the traditional (Band) component at $\la 1\;$MeV and the UP emission to the high-energy emission observed
by {\it Fermi}/LAT for some GRBs at $\ga 10\;$MeV. The two components can be separate in the spectrum 
in some cases or can mimic a smooth broad Band spectrum in other cases. We apply our formulation to the 
well-studied long and short LAT GRBs, GRB 080916C, GRB 090902B, and GRB 090510, and typically
find reasonable parameters for fitting the time-binned spectra, although fine tuning of several 
parameters is required. The observed delays of the high-energy emission 
with respect to the MeV emission which are large compared to the variability times are unlikely 
to be due to simple kinematic effects of a non-evolving jet. These delays may instead be attributed to the 
temporal evolution of the physical parameters of the jet, and thus the delay timescales could 
provide a potential tool for investigating the structures of GRB jets themselves and 
their progenitors. The difference of the delay timescales of long and short GRBs inferred from 
the {\it Fermi} data might be due to the differences in the progenitors of long and short GRBs.
Some other properties and consequences of this model are discussed, including 
temporal correlations among the prompt optical, the soft X-ray, and the distinct 
high-energy component as well as the Band component.
\end{abstract}


\section{Introduction}
\label{sec:intro}

The origin of gamma-ray burst (GRB) prompt emission remains unclear. 
It has been mainly observed in the energy range of $10\;{\rm keV} - 1\;{\rm MeV}$,
and most of the spectra are fitted by a simple broken power-law function (so-called Band function)
or a cutoff power-law function \citep{preece00,ghirlanda02,kaneko06}, the light curves being typically 
very variable. Many models have been proposed to explain these properties.
Most of them invoke a relativistic jet that is energized by a newly born compact object. 
Currently three types of emission mechanisms in the relativistic jet are being actively discussed;
photospheric, leptonic synchrotron, and hadronic emission models. The first models assume that 
the thermal energy stored in the jet can be radiated as prompt emission at the Thomson photosphere
\citep[e.g.,][]{paczynski86,goodman86,thompson94,meszaros00}. Here the thermal 
energy in the jet can be produced by the particle and/or magnetic energy dissipation between
the explosion center and the photosphere. The second and third models assume the Thomson-thin 
region as the prompt emission site. In the second models, the electrons (and positrons) 
accelerated by the shock dissipation of the kinetic energy or by the magnetic energy dissipation
radiate synchrotron and synchrotron-self-Compton (SSC) emissions
\citep[e.g.,][]{rees94,daigne98,kumar08,spruit01,lyutikov06}, while in the third models the accelerated
protons and secondary particles induced by the photopion cascade process produce synchrotron and
inverse Compton (IC) emissions \citep[e.g.,][]{vietri97,bottcher98,asano09b}.
Clarifying the prompt emission mechanism and the physical quantities at the emission site would
help understand the nature of the relativistic jet and the compact object that energizes the jet.
For other models and more general discussions, see recent reviews 
\citet{piran04}, \citet{meszaros06}, and \citet{zhang07}.

GRBs were only sparsely observed in the $> 10\;$MeV energy range, until the {\it Fermi} satellite
was launched 2008 June 11. Now the GBM ($8\;{\rm keV} - 40\;{\rm MeV}$) and the LAT ($\sim 20\;{\rm MeV}
- 300\;{\rm GeV}$) detectors onboard {\it Fermi} provide extremely broad energy coverage with good
temporal resolution for GRBs. The {\it Fermi} observations will put further constraints on the 
above three types of prompt emission models. During its first $1.5\;$yr routine operation, the LAT 
has detected 14 GRBs. Those are summarized in \citet{granot10}. Most of their spectra are fitted
by a Band function even up to $\sim 10\;$GeV, 
while at least 3 GRBs (GRB 090510, GRB 090902B, and GRB 090926A) have additional distinct 
spectral component at $\ga 10\;$MeV. Those additional components are fitted by single power-law function. 
These 3 GRBs are among the brightest GRBs detected by {\it Fermi}. This suggests that such a
high-energy component may be very common and we can clearly detect such a distinct high-energy component 
only in bright LAT GRBs \citep{granot10}.

{\it Fermi} also revealed that the high-energy emission ($> 100\;$MeV) of most LAT GRBs is delayed
behind the onset of the MeV emission, and the high-energy emission of many LAT GRBs lasts longer
than the MeV emission, showing power-law decays, which are typically detected until 
$\sim 10^2 - 10^3\;$s. The delay times in the cosmological rest frame
are $\sim 1\;$s for long GRBs and $\la 0.1\;$s for short bursts GRB 081024B
and GRB 090510. Although some delays of the high-energy photons are just caused by the flux 
increases above the LAT detection threshold without a spectral change, others must clearly be attributed to 
the spectral changes of the Band component and/or the onset of the distinct spectral component. 
We note that the distinct components of the 3 GRBs are delayed.

The origins of the distinct spectral components and the onset delays in the high-energy range
have been actively debated. It has been proposed that the high-energy emission can be attributed to 
the external shock, which is made by the interaction of the jet with the ambient medium, and 
the onset delays can correspond to the times for the jet decelerations 
\citep{kumar09,ghisellini10} \citep[see also][]{granot03,peer04}. However, the observed 
high-energy emission usually has a strong variability and it often correlates with 
the MeV emission, which is at odds with an external shock origin.
Especially for GRB 090510, whose long-lived high-energy emission can be explained by the 
external shock synchrotron emission together with the optical and X-ray afterglows detected from
$\sim 100\;$s after the burst trigger \citep{depasquale10,kumar10,corsi10}, it is found that the 
high-energy emission in the prompt phase is much brighter than that produced by the same external 
shock \citep{he10}.  A similar conclusion has been obtained for GRB 090902B \citep{liu10}.
As for the internal shock models, it is not simple to explain the LAT onset delays which are 
larger than the variability timescales ($t_v$) by the delayed brightening of the SSC emission 
\citep{abdo080916C,abdo090902B,ackermann090510,corsi10,daigne10}. 
The spectral index of the distinct high-energy 
component different from the Band low-energy spectral index, seen in e.g., GRB 090902B, may 
not be explained simply either in these models. To overcome these problems, we have
proposed an external inverse Compton (EIC) component (i.e., the emission produced by 
up-scattering the photons incident from outside the shock) in addition to the synchrotron 
and SSC components in the internal shock model \citep{toma09}; this model, however, 
requires an extreme value of the microphysical parameter $\epsilon_B \sim 10^{-5}$, which
does not seem common.
Other alternative mechanisms such as hadronic emission mechanisms require huge isotropic energies
\citep{asano09,wang09,razzaque10}, while magnetically-dominated jets 
have not been explored explicitly enough for detailed comparisons 
with {\it Fermi}/LAT observational results \citep[see][]{zhang09,fan10,zhang10}.

In this paper we concentrate on a photospheric emission model. Such models
have been briefly discussed for interpreting the relation of the MeV emission with 
the high-energy emission. In this scenario, the photosphere produces a Band emission 
component which peaks around $\sim 1\;$MeV, and this cannot generally produce the
high-energy emission because of large opacity for $e^{\pm}$ pair creation. The high-energy 
emission in these models may instead arise in a dissipation region at a larger radius 
\citep{gao09,beloborodov10,ryde10}.  In this paper, we substantially extend this idea 
and discuss its consequences in greater depth.  We focus on the case 
in which the energy of the jet is mainly carried by photons and baryons (where magnetic field 
energy is subdominant), and the dissipation 
mechanism at large radius is internal shock. We show that in typical cases the photospheric 
emission is efficiently Compton scattered by the electrons in internal shocks outside the photosphere,
and find that the up-scattered photospheric (UP) emission is a good candidate for 
the observed high-energy emission in the prompt phase. This Compton scattering is a type of the
EIC scattering, which is thought to be important for the high-energy emission of blazars 
\citep[e.g.,][]{sikola94,dermer93,brunetti00,ghisellini09}, and could also be important in the
internal and external shocks in GRBs \citep{beloborodov05,wang06,fan08,toma09,murase10,murase10b}.
We make a generic formulation of the spectral types of the radiation from the photosphere
and the internal shock including synchrotron and SSC emission for the cases of the efficient
EIC scattering, and clarify necessary conditions for the photospheric and UP components
to be dominant in the energy range of $10\;{\rm keV} - 10\;{\rm GeV}$, rather than synchrotron
and SSC components. This formulation is applied for the well-observed LAT long and short GRBs,
GRB 080916C, GRB 090902B, and GRB 090510, and we typically find reasonable parameter sets for 
which the data can be explained by the photospheric and UP emission. Our model fits indicate 
that the observed delayed onset of the LAT emission may be interpreted as the parameter evolution 
of the GRB jet.

The UP emission ceases at the end of the prompt MeV (photospheric) emission,
and the subsequent long-lived high-energy emission may instead be related to the external shock 
\citep[e.g.,][]{depasquale10,kumar10,ghisellini10}.  A different origin for the prompt and late LAT
emission is in fact implied by the change in its behavior across this transition \citep{he10,liu10}.
In what follows, we concentrate on the prompt emission phase.
We discuss the general temporal and spectral properties of the radiation 
from the photosphere and the internal shock of
the GRB jet in Sections \ref{sec:ph}, \ref{sec:temporal}, and \ref{sec:spectral}, and perform
the case studies of GRB 080916C, GRB 090902B, and GRB 090510 in Section \ref{sec:case}.
Some of the implications of our results are discussed in Section~\ref{sec:summary}.

\section{Photospheric emission}
\label{sec:ph}

We consider that the jet is accelerated by the thermal pressure, and the magnetic field energy 
is subdominant. We derive the luminosity and temperature of the photospheric
emission from the jet and the remaining kinetic luminosity of the jet above the photosphere,
which is the luminosity budget for internal shocks, according to the standard fireball model
\citep{paczynski86,paczynski90,shemi90,meszaros93,meszaros00,nakar05}. We assume that
the jet material is optically thick and the energy is dominated by thermal 
radiation at the base, $r=r_a$, from where it expands in the roughly adiabatic condition,
i.e., without strong conversion of the kinetic energy into the thermal energy affecting 
the adiabatic expansion dynamics \citep[for a continuous strong dissipation case, see][]{rees05}. 
The jet material may have a Lorentz factor $\Gamma_a \geq 1$
at the base. 
In particular, if the GRB originates from a massive stellar collapse and the jet suffers strong 
dissipation of its kinetic energy due to interaction with the stellar
envelope, the front portion of the jet may have $r_a \sim 10^{10} - 10^{11}\;$cm and $\Gamma_a \sim
10 - 10^2$ \citep[see recent numerical simulations by][and discussion in 
Section~\ref{sec:summary}]{morsony07,lazzati09}. 
In the absence of a strong dissipation the base radius may be as small as a few of Schwarzschild radius
of the central compact object, so that $r_a \sim 10^6 (M_c/M_{\odot})\;$cm and $\Gamma_a \sim 1$,
where $M_c$ is the mass of the central object.
 
The observer-frame temperature at the base is
\begin{equation}
T_a = \left(\frac{L \Gamma_a^2}{4\pi r_a^2 c a}\right)^{1/4}
\simeq 2\; L_{53}^{1/4} \left(\frac{r_{a,7}}{\Gamma_a}\right)^{-1/2}\; {\rm MeV}/k,
\label{eq:T_a}
\end{equation}
where $L$ is the isotropic equivalent luminosity of the jet, and $a \simeq 7.56 \times 10^{-15}\;
{\rm erg}\;{\rm cm}^{-3}\;{\rm K}^{-4}$ is the Stefan constant.\footnote{We adopt the notation 
$Q_x = Q/10^x$ in cgs units throughout this paper.} For general cases of GRBs,
the adiabatic index of the jet material below the photosphere is found to be 4/3 (see below).
The dynamics of the expanding jet can be summarized as $\Gamma \propto r$ and $T = T_a$ for
$r \leq r_s$, and $\Gamma = \eta$ and $T \propto r^{-2/3}$ for $r \geq r_s$, where the Lorentz
factor saturation radius $r_s = r_a \eta / \Gamma_a$. Here we have defined $\eta \equiv L/\dot{M}c^2$, 
where $\dot{M}$ is the isotropic equivalent mass ejection rate. We see that the ratio of the 
radiation and baryon entropy densities $\sim a {T'}^4/n' kT'$ is constant throughout the evolution 
below the photosphere and is much larger than unity if $\eta \gg k T_a/(m_p c^2) \simeq 
2 \times 10^{-3}\;L_{53}^{1/4} (r_{a,7}/\Gamma_a)^{-1/2}$ 
(where $n'$ and $T'$ is the baryon density and fluid temperature in the comoving frame).
Thus the assumption of the adiabatic index $= 4/3$ is validated.

The $e^{\pm}$ pairs drop out of equilibrium at $T' \sim 20\;$keV, at a radius $r \simeq r_p
=1 \times 10^{9}\; L_{53}^{1/4} (r_{a,7}/\Gamma_a)^{1/2}\;$cm. We consider the case in which
the jet carries enough baryons to provide an electron scattering photosphere above $r_p$
(i.e., the case of $\eta < 5 \times 10^6\;L_{53}^{1/4} (r_{a,7}/\Gamma_a)^{1/2}$).
Then the photosphere radius $r_{\rm ph}$ is defined as $\tau = \sigma_T n' r/(2\Gamma) = 1$, where
$\sigma_T$ is the Thomson cross section and the comoving electron density is given by
$n' = L/(4\pi r^2 m_p c^3 \eta \Gamma)$, if there is no additional $e^{\pm}$ pair creation 
in the jet \citep{abramowicz91}.

If no energy dissipation takes place around the photosphere, the photons and particles are 
fully thermalized, typically with the comoving temperature $T'_{\rm ph} \sim 0.1 - 1\;$keV, and then
the emerging radiation from the photosphere is a blackbody with the observer-frame temperature 
$T_{\rm ph} \simeq \Gamma_{\rm ph} T'_{\rm ph}$ (where $\Gamma_{\rm ph}$ is the Lorentz factor 
at $r = r_{\rm ph}$).
However, some energy dissipation processes are expected to occur around the photosphere,
which can make the emerging radiation spectrum deviate from a blackbody.
Such processes have been demonstrated convincingly by recent analytical and numerical work:
Both below and above the photosphere, a fraction of the electrons can acquire high-temperature 
distribution ($\gg T'_{\rm ph}$), either due to internal shocks and/or interaction of a jet with 
stellar envelope \citep{eichler00,rees05,peer05,thompson07,lazzati10}, 
dissipation of excited plasma waves \citep{ioka07},
or nuclear collisions between protons and neutrons \citep{beloborodov10}, or in magnetically-dominated
jets due to scattering with turbulent Alfv\'{e}n waves \citep{thompson94} or heating caused
by magnetic reconnection \citep{giannios06}. Multiple IC scatterings of the thermal photons 
by mildly relativistic electrons (with $\gamma \sim 1$) can create a power-law tail extending
from the thermal peak, which saturates at the comoving-frame energy $\varepsilon' \sim m_e c^2$
as the Klein-Nishina limit or by the direct Compton cooling, which is boosted to $\varepsilon \sim
0.1-1\;$GeV in the observer frame. The dissipation processes also can produce a relativistic
electron population, which IC-scatter the photons to even higher energies, the limitation being the 
$\gamma\gamma$ absorption against lower energy photons. 
The low-energy spectral slope of the emerging photons should not be significantly changed in such
processes, i.e., the photon index $\alpha_{\rm ph} \simeq 1$ \citep{beloborodov10,lazzati10}.

In our baryonic jet model, we assume that the dissipation of the kinetic energy is not so strong that 
the jet dynamics is described by the adiabatic evolution, and the temperature of the emerging
photons is given by $T_{\rm ph}$. This implies that the energy flux in the possible non-thermal
part above $\varepsilon_{\rm ph}$ should be smaller than that in the main thermal part, 
leading to an upper limit on the photon index $\beta_{\rm ph}$ of the possible power-law tail.
Let us simply write down the spectral shapes of the pure thermal component and the thermal plus a 
non-thermal tail spectrum as $F_{\varepsilon}^{\rm th} = 0.84
(\varepsilon/k T_{\rm ph})^3/[\exp({\varepsilon/k T_{\rm ph}})-1]$ and $F_{\varepsilon}^{\rm nth}
= (\varepsilon/\varepsilon_{\rm ph})^2 {\min}[1, (\varepsilon/\varepsilon_{\rm ph})^{\beta_{\rm ph}-1}]$,
respectively, where $T_{\rm ph} = \varepsilon_{\rm ph}/4$, and these are normalized as
$F_{\varepsilon}^{\rm th} = F_{\varepsilon}^{\rm nth}$ at $\varepsilon = \varepsilon_{\rm ph}$.
\footnote{
More accurately, one should use $F_{\varepsilon}^{\rm th} = 0.7
(\varepsilon/k T_{\rm ph})^3/[\exp({\varepsilon/k T_{\rm ph}})-1]$ where $T_{\rm ph} = 
\varepsilon_{\rm ph}/3$ for the pure thermal component since $\varepsilon = 4 kT_{\rm ph}$ is 
the peak energy of the $\varepsilon F_{\varepsilon}$ spectrum instead of the $F_{\varepsilon}$ spectrum.
However, we obtain the same constraint $\beta_{\rm ph} \la -2.5$ also from this calculation. 
}
Then we have $\int^{\infty}_0 F_{\varepsilon}^{\rm th} d\varepsilon = 0.84(\pi^4/60)\varepsilon_{\rm ph}$
and $\int^{\infty}_0 F_{\varepsilon}^{\rm nth} d\varepsilon = 
[(1/3)-(\beta_{\rm ph}+2)^{-1}]\varepsilon_{\rm ph}$.
The condition $\int^{\infty}_0 F_{\varepsilon}^{\rm nth} d\varepsilon < 2 \times
\int^{\infty}_0 F_{\varepsilon}^{\rm th} d\varepsilon$ provides a constraint $\beta_{\rm ph} \la -2.5$.
The value of $\beta_{\rm ph}$ is expected to be determined not only by the ratio of the photon to 
kinetic energies but also on the electron energy distribution which depends on the type of a dominant 
dissipation process, which we may not specify at the current stage. Hereafter we treat $\beta_{\rm ph}$ 
as a free parameter ranging within the above constraint.

In such dissipation processes copious $e^+e^-$ pairs may be created, which increase
the Thomson photosphere radius. We parameterize the number density 
of electrons plus positrons (which we will call `leptons' hereafter) near the photosphere as 
$n'_l = \mathcal{R} n'$, where $\mathcal{R} \geq 1$. Typically we can have $\mathcal{R} \sim 10 - 10^2$, 
in which case the inertia is still dominated by baryons. Thus the photosphere is defined as $\tau_l = 
\sigma_T n'_l r/(2\Gamma) = 1$.

We have $r_{\rm ph} < r_s$ in the low baryon load case $\eta > \eta_*$, where
\begin{equation}
\eta_* = \left(\frac{\sigma_T \mathcal{R} L \Gamma_a}{8\pi r_a m_p c^3}\right)^{1/4}
\simeq 2.8 \times 10^3\; L_{53}^{1/4} \left(\frac{r_{a,7}}{\Gamma_a}\right)^{-1/4} \mathcal{R}_1^{1/4}.
\label{eq:eta_*}
\end{equation}
In this case, most of the luminosity is radiated at the photosphere. Leptons can frequently interact
with photons even above the photosphere, and this Compton drag acceleration determines the remaining
kinetic luminosity \citep{meszaros93}.
The emission radius, luminosity, and peak energy of the photospheric emission and the final bulk Lorentz
factor and remaining kinetic luminosity of the jet above the photosphere are given by \citep{meszaros00}
\begin{eqnarray}
&& r_{\rm ph} = r_a \frac{\eta_*}{\Gamma_a} \left(\frac{\eta}{\eta_*}\right)^{-1/3}, ~~~
L_{\rm ph} \approx L, ~~~
\varepsilon_{\rm ph} \simeq 4\; k T_a, ~~~ \nonumber \\
&& \Gamma_f \simeq \eta_*, ~~~
L_k \simeq L \left(\frac{\eta}{\eta_*}\right)^{-1}.
\label{eq:ph_low_baryon}
\end{eqnarray}
On the other hand, in the high baryon load case $\eta < \eta_*$, we have $r_{\rm ph} > r_s$.
In this case most of the initial thermal energy has been converted to the kinetic energy below
the photosphere. The corresponding quantities are given by\footnote{We do not discuss the 
case $r_{\rm ph} > r_{sp} = 2 W \eta^2$, for which the radial 
spreading of shells (with the initial width $W$) is significant, and the scaling law is different, 
$r_{\rm ph} \propto \eta^{-1/2}$ \citep{nakar05}.}
\begin{eqnarray}
&& r_{\rm ph} = r_a \frac{\eta_*}{\Gamma_a} \left(\frac{\eta}{\eta_*}\right)^{-3}, ~~~
L_{\rm ph} \simeq L \left(\frac{\eta}{\eta_*}\right)^{8/3}, ~~~ \nonumber \\
&& \varepsilon_{\rm ph} \simeq 4\; k T_a \left(\frac{\eta}{\eta_*}\right)^{8/3}, ~~~
\Gamma_f = \eta, ~~~
L_k \approx L. 
\label{eq:ph_high_baryon}
\end{eqnarray}
For typical GRB jets with parameters $L \sim 10^{51} - 10^{54}\;{\rm erg}\;{\rm s}^{-1}$, 
the photospheric emission can be bright with $\varepsilon_{\rm ph} \sim 100\;{\rm keV} - 10\;{\rm MeV}$
since the jets can have $10^{-1} \la r_{a,7}/\Gamma_a \la 10^2$ and $\eta \ga \eta_*$ in principle.

\section{Temporal Properties}
\label{sec:temporal}

\begin{figure*}
\begin{center}
\includegraphics{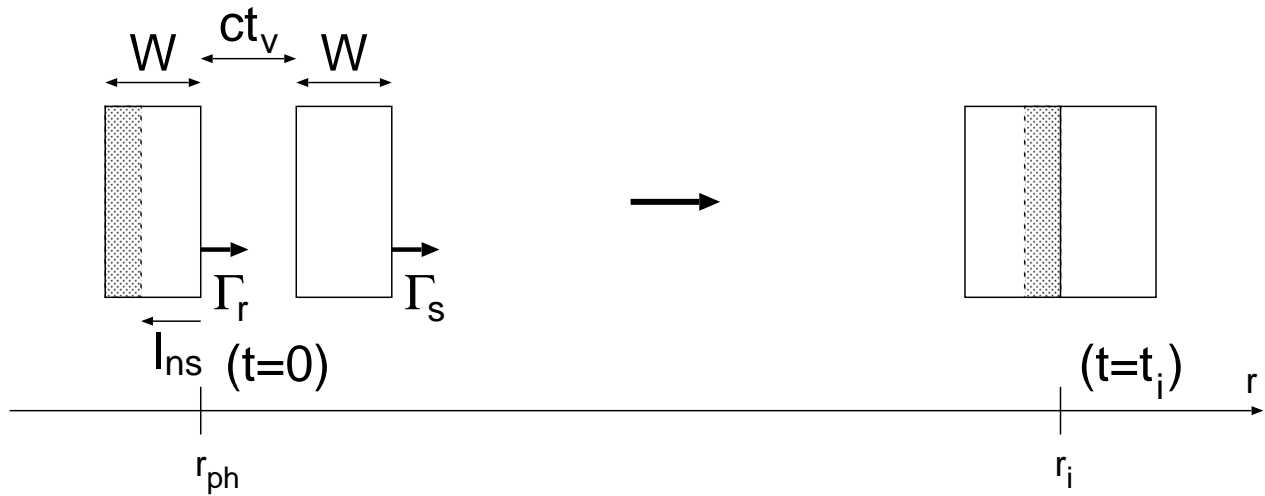}
\end{center}
\caption{
Interaction of the photospheric emission from two relativistically moving shells with the 
internal shock that the given shells themselves give rise to. The photons from the shaded region can 
be scattered by the leptons in the internal shock.
}      
\label{fig:is_scat}
\end{figure*}

The GRB jet may be thought to consist of many successive shells with initial radial widths
$W \ll r_{\rm ph}$. This implies that the photospheric emission is temporally variable.
After emerging from the photospheric regions, collisions between the ejected shells can
produce internal shocks at $r_i \gg r_{\rm ph}$. As we will show below, the leptons in the 
internal shock of two given shells can up-scatter their own photospheric emission, which was 
produced as they emerged from the photosphere. The UP emission can appear as a high-energy spectral
component in the LAT energy range (see Section~\ref{sec:spectral}). The temporal properties of the emission
in this model can be exemplified using a simple two-shell collision model 
(see Figure~\ref{fig:is_scat}). For simplicity, we only consider the cases in which
the two shells are both in the regime $\eta < \eta_*$ or both in the regime $\eta > \eta_*$.

The duration of the observed photospheric emission from one shell is given by the 
light crossing time of the shell plus the angular spreading time,
\begin{equation}
\delta t_{\rm ph} \sim \frac{W}{c} + \frac{r_{\rm ph}}{2 c \Gamma^2}.
\end{equation}
This can be rewritten as $\delta t_{\rm ph} \sim (W/c) + r_{\rm ph}/(2c\eta^2)
\simeq (W/c) + (r_{\rm ph}/r_i)t_v$ for $\eta < \eta_*$,
and $\delta t_{\rm ph} \sim (W/c) + r_{\rm ph}/(2c\Gamma_{\rm ph}^2)
\simeq (W/c) + (r_f/r_i)(\eta/\eta_*)^{1/3}t_v$ for $\eta > \eta_*$,
where $\Gamma_{\rm ph} = \eta_*(\eta/\eta_*)^{-1/3}$, and 
$r_f = r_a \eta_*/\Gamma_a$ is the radius at which the Compton drag acceleration ceases.

Now we consider a two-shell collision. We assume that the two shells have similar initial
radial width $W$, and the initial separation is $c t_v$. First consider the high baryon
load case $\eta < \eta_*$. The rapid shell with Lorentz factor $\Gamma_r = \eta_r$ catches up 
with the slower shell with Lorentz factor $\Gamma_s = \eta_s$ (Hereafter subscripts `r' and `s'
denote the quantities of the rapid and slow shells, respectively). Setting the zero time in the 
lab frame as the time when the rapid shell is at $r = r_{\rm ph}$, the collision time is given 
by $t_i = ct_v/[c(\beta_r - \beta_s)] \approx 2 c t_v \eta_s^2$, where we have used an approximation 
$\eta \gg 1$. 
The dissipation of the total kinetic energy is most efficient when the masses of the two shells
are similar \citep[c.f.,][]{piran04}. For typical cases, the ratio of the two Lorentz factors may be 
$a_L = \eta_r/\eta_s \la 5$. Then the dissipation efficiency is given by $\epsilon_d \la 0.25$.

The photons which arrive at $r=r_i$ at time $t \geq t_i$ can be up-scattered by the energetic
leptons in the internal shock. Let us define a length $l_{\rm ns}$, measured from the front end
of the rapid shell, as the radial scale of the region from which the photospheric photons do 
not interact with the leptons in the internal shock. Then we have $r_{\rm ph} - l_{\rm ns} + ct_i
= r_i = r_{\rm ph} + c\beta_r t_i$, which gives
\begin{equation}
l_{\rm ns} = c(1-\beta_r)t_i \approx a_L^{-2} ct_v. ~~~~~~(\eta < \eta_*)
\label{eq:lns}
\end{equation}
We find that inverse Compton scattering is efficient for the case $l_{\rm ns} < W/2$ and 
inefficient for the case $l_{\rm ns} \ga W/2$.

A similar argument can be made also for the low baryon load case $\eta > \eta_*$. In this case
the shells are accelerated similarly and cannot collide with each other below 
$r_f = r_{\rm ph} (\eta/\eta_*)^{1/3}$.
The internal shock radius is given by $r_i \approx 2 c t_v \eta_{*,s}^2$, and we consider a case
$r_i \gg r_{f,r}$. We have 
$l_{\rm ns} = ct_i - \int^{t_i}_0 c \beta_r dt \approx (\eta_{*,s}^2/\eta_{*,r}^2) c t_v 
+ (r_{{\rm ph},r} \Gamma_{{\rm ph},r}^{-2} - r_{f,r} \eta_{*,r}^{-2})$. The second term represents
a correction due to the velocity of the shell lower than $\eta_{*,r}$ at $r < r_{f,r}$.
The second term is rewritten as $r_{f,r} \eta_{*,r}^{-2} [(\eta_r/\eta_{*,r})^{1/3} -1]$,
which cannot be dominant unless $\eta \gg \eta_*$, so that $l_{\rm ns}$ can typically be written as 
Equation~(\ref{eq:lns}) also in this case.

We have two parameter regimes for the scattering efficiency:

{\it (i) Efficient scattering regime,} $l_{\rm ns} < W/2$. This includes the typical case $W \sim c t_v$.
In this regime we may observe bright UP emission. The duration of the photospheric emission
can be written as $\delta t_{\rm ph} \sim W/c$, since we have assumed that $r_i \gg r_{\rm ph}$
for $\eta < \eta_*$ and $r_i \gg r_f$ for $\eta > \eta_*$. The onset of the UP emission is correlated 
with that of the photospheric emission pulse released from the rapid shell and delayed from that of
the photospheric emission pulse released from the slow shell by 
$t_{\rm lag} = (W + ct_v + l_{\rm ns})/c \sim (W/c) + t_v$. 
The duration of the UP emission is given by the duration of the seed photons plus the angular 
spreading time 
\begin{equation}
\delta t_{\rm up} \sim \frac{W - l_{\rm ns}}{c} + \frac{3 r_i}{2c\Gamma_m^2} \simeq 
\frac{W}{c} + 3 a_L^{-1} t_v.
\label{eq:up_dur}
\end{equation}
The factor of $\sim 3$ in the second term means that the UP emission has an anisotropic energy distribution 
in the comoving frame, being brightest at an angle $\theta \sim 1/\Gamma_m$ from the line of sight 
in the observer frame \citep[c.f.,][]{brunetti00,wang06,fan08,toma09}, where 
$\Gamma_m \simeq (\Gamma_r \Gamma_s)^{1/2}$ is the Lorentz factor of the merged shell.

{\it (ii) Inefficient scattering regime,} $l_{\rm ns} \ga W/2$. In this regime, typically we may not
observe bright UP emission. A caveat is that this regime includes a case $l_{\rm ns} \sim W + c \tilde{t}_v$,
in which we have a third shell behind the rapid shell of the two given shells apart by 
$c \tilde{t}_v$, and the photospheric emission from the third shell can be scattered by the leptons 
in the internal shock of the two given shells. This condition reduces to
$a_L^{-2} c t_v \sim W + c \tilde{t}_v$, which is satisfied when $\tilde{t}_v \ll t_v$.

If the GRB jet is in the efficient scattering regime right from the start, when it first
emerges, the delay timescale between the onsets of the first photospheric emission (in the MeV
energy range) and the first UP emission (in the high-energy range) is $\sim t_{\rm lag} 
\sim (W/c) + t_v$, which is comparable to $\delta t_{\rm ph}$ or $t_v$.
However, {\it Fermi} observations show that the LAT onset delays are much larger than the variability 
timescale apparent in the MeV energy light curves both for long and short GRBs. 
Therefore it is unlikely that 
the large delays of the high-energy emission onsets observed in many GRBs are due to the above 
simple kinematic effect of the jet whose physical parameters do not evolve. In our model, the large 
delays may be interpreted as the timescale on which the physical parameters of the jet temporally
change, e.g., $W$ and $t_v$ change from the inefficient scattering regime into the efficient
scattering regime, or $L$, $r_a/\Gamma_a$, $\eta$, and $\mathcal{R}$ change from the regime 
$\eta > \eta_*$ into the regime $\eta < \eta_*$ (see the following sections for details of
the latter possibility). In these cases the UP emission can start, being
still dim compared with the photospheric emission, at a time $\sim t_{\rm lag}$ after the first 
photospheric emission, and can become bright with the delay as observed.

\section{Spectral Properties}
\label{sec:spectral}

As shown above, the photospheric emission of the shells can be efficiently up-scattered by the 
leptons in the internal shocks of the same shells. Here we derive the generic broadband spectrum 
of the observed radiation arising in the internal shock including synchrotron and SSC emission for 
the efficient scattering regime.

\subsection{Case of $\eta < \eta_*$}

In this case the kinetic luminosity that can be dissipated into radiation
by the internal shock is $L_k \approx L$.
At $r > r_{\rm ph}$, we have $\sigma_T n'_l r/(2\eta) < 1$, and the pair annihilation timescale
is longer than the expansion timescale. Thus the pair population freezes out and we can write
$n'_l = \mathcal{R} L/(4\pi r^2 m_p c^3 \eta^2)$ even well above the photosphere. 
Hereafter we focus on the photospheric emission from the rapid shell, which is up-scattered
by the leptons in the internal shock, so that $\eta$ denotes $\eta_r$ (The photospheric
emission from the slower shell may be much dimmer).
The internal shock radius is estimated as $r_i \simeq 2 c t_v \eta_s^2 = 2 c t_v \eta^2 a_L^{-2}
\simeq 2 \times 10^{13} \eta_3^2 t_{v,-2} (a_L/5)^{-2}\;$cm.
Our assumption $r_i \gg r_{\rm ph}$ reduces to 
\begin{equation}
\eta \gg 5 \times 10^2\;L_{53}^{1/5} \mathcal{R}_1^{1/5} t_{v,-2}^{-1/5} \left(\frac{a_L}{5}\right)^{2/5}.
\label{eq:cond_high}
\end{equation}

We assume that all the leptons participate in the non-thermal acceleration process in the internal shock
and have the injected spectrum $dn'_{l,i}/d\gamma = C \gamma^{-p}$ for $\gamma \geq \gamma_m$, 
where $C$ is a constant. Then we have $C \gamma_m^{1-p}/(p-1) = n'_l(r=r_i)$ and 
$C m_e c^2 \gamma_m^{2-p}/(p-2) = L\epsilon_d \epsilon_e / (4\pi r_i^2 c \eta^2)$,
where $\epsilon_e$ is the fraction of the dissipation energy that is carried by leptons
in the internal shock.
(We consider a relatively weak internal shock, i.e., $a_L \la 5$, where we can neglect
the change of the Lorentz factor of the jet for simplicity.) These lead to
\begin{equation}
\gamma_m = \frac{m_p}{m_e} \frac{p-2}{p-1} \mathcal{R}^{-1} \epsilon_d \epsilon_e
\simeq 4\; \mathcal{R}_1^{-1} \left(\frac{\epsilon_d \epsilon_e}{0.1}\right) f(p),
\label{eq:gamma_m}
\end{equation} 
where $f(p) = 13(p-2)/[3(p-1)]$. If only a fraction of leptons are accelerated, $\gamma_m$
is larger than this value.

We also assume that a fraction $\epsilon_B$ of the dissipated energy is carried in the form
of magnetic fields, $U'_B = {B'}^2/(8\pi) = L \epsilon_d \epsilon_B/(4\pi r_i^2 c \eta^2)$.
The characteristic synchrotron energy of leptons with Lorentz factor $\gamma$ is given by
\begin{eqnarray}
&& \varepsilon_{\rm syn}(\gamma) \simeq \frac{3 h e B'}{4\pi m_e c}\gamma^2 \eta \nonumber \\
&& \simeq 6\times 10^{-1}\; \gamma^2 \; L_{53}^{1/2} \eta_{3}^{-2} t_{v,-2}^{-1} \left(\frac{a_L}{5}\right)^2
\left(\frac{\epsilon_d \epsilon_B}{0.1}\right)^{1/2}\; {\rm eV}.
\end{eqnarray}

The leptons emit the UP emission, synchrotron emission, and SSC emission. The comoving energy 
density of the synchrotron emission can be written by
\begin{equation}
U'_{\rm syn} = t'_{\rm dyn} \frac{4}{3} \sigma_T c U'_B \int \gamma^2 \frac{dn'_l}{d\gamma}
d\gamma = x U'_B,
\end{equation}
where $t'_{\rm dyn} \simeq r_i/(2c\eta)$ is the dynamical timescale of the internal shock 
and $dn'_l/d\gamma$ is the lepton energy distribution averaged over the dynamical timescale.
The quantity $x$ is calculated as
\begin{equation}
x = \frac{4}{3} \sigma_T \frac{r_i}{2\eta} \int \gamma^2 \frac{dn'_l}{d\gamma} d\gamma
\simeq \frac{4(p-1)}{3(p-2)} \tau_{l,i} \gamma_m \gamma_c h(\gamma_m,\gamma_c),
\label{eq:x}
\end{equation}
where $\gamma_c$ is the cooling Lorentz factor of the leptons and $\tau_{l,i}$ is the Thomson
optical depth at $r_i$ (which can be written as $\tau_{l,i} = r_{\rm ph}/r_i$ in the case of $\eta < \eta_*$). 
We have assumed $p>2$ and defined
the function $h(\gamma_m,\gamma_c)$ as $h=1$ for $\gamma_c \ll \gamma_m$, $h=p/(p-1)$ for 
$\gamma_c \approx \gamma_m$, $h=(\gamma_c/\gamma_m)^{2-p}/(3-p)$ for $\gamma_m \ll \gamma_c$
and $p<3$, and $h = (p-2)\gamma_m/[(p-3)\gamma_c]$ for $\gamma_m \ll \gamma_c$ and $p \geq 3$.
Similarly we have the SSC emission energy density $U'_{\rm ssc} = x U'_{\rm syn}$, where
we have assumed that the Klein-Nishina (KN) effect is not significant. This is valid for
the parameters adopted in the case study of observed LAT GRBs in Section~\ref{sec:case}. 
We also have the UP emission
energy density $U'_{\rm up} = x U'_{\rm ph}$, where we have assumed that the KN effect is not 
significant also for the UP emission. This is found to be valid for observed LAT GRBs 
in Section~\ref{sec:case}. We also have neglected the anisotropy of the UP 
photon energy distribution, for simplicity. This anisotropy leads to the reduction of the 
observed UP luminosity averaged over a pulse than the isotropic assumption by a factor of $\sim 2$
\citep{fan08,toma09}, which we will neglect below. By using the relation, e.g., 
$L_{\rm ph} = 4\pi r_i^2 c \eta^2 U'_{\rm ph}$, we have
\begin{eqnarray}
L_{\rm up} &\simeq& x L_{\rm ph}, \nonumber \\
L_{\rm syn} &\simeq& \epsilon_d \epsilon_B x L \simeq k x L_{\rm ph}, \\
L_{\rm ssc} &\simeq& \epsilon_d \epsilon_B x^2 L \simeq k x^2 L_{\rm ph} \nonumber,
\end{eqnarray}
where we have defined
\begin{equation}
k \equiv \frac{\epsilon_d \epsilon_B}{(\eta/\eta_*)^{8/3}}.
\end{equation}
The photospheric luminosity is given by $L_{\rm ph} \simeq L (\eta/\eta_*)^{8/3}$ 
(see Eq.~\ref{eq:ph_high_baryon}).

\begin{table*}
\centering
\caption{Ordering of emission luminosities for various cases of $\mathcal{G} = (\eta/\eta_*)^{8/3}$,
$\mathcal{E} = \epsilon_d \epsilon_e h$, and $\mathcal{B} = \epsilon_d \epsilon_B$.}
\begin{tabular}{ccllcc}
\hline
Case  &  $\mathcal{G}$ & $k ({\rm or} ~k'), x$ & & Luminosities & $\mathcal{E}$ and $\mathcal{B}$ \\
\hline 
\hline
1 & $\eta < \eta_* (\mathcal{G}<1)$ & 
$k \ll 1, ~kx \ll 1$ & $x \ll 1$ & $L_{\rm ph} \gg L_{\rm up} \gg L_{\rm syn} \gg L_{\rm ssc}$ &
$\mathcal{G} \gg {\rm max}(\mathcal{E},\mathcal{B})$ \\ 
2 & & & $x \gg 1, ~kx^2 \ll 1$ &
$L_{\rm up} \gg L_{\rm ph} \gg L_{\rm ssc} \gg L_{\rm syn}$ &
$\mathcal{E} \gg \mathcal{G} \gg (\mathcal{E}^2 \mathcal{B})^{1/3} \gg \mathcal{B}$ \\ 
3 & & & $x \gg 1, ~kx^2 \gg 1$ &
$L_{\rm up} \gg L_{\rm ssc} \gg L_{\rm ph} \gg L_{\rm syn}$ &
$\mathcal{E} \gg (\mathcal{E}^2 \mathcal{B})^{1/3} \gg \mathcal{G} \gg \mathcal{B}$ \\
4 & & $k \ll 1, ~kx \gg 1$ & & $L_{\rm ssc} \gg L_{\rm up} \gg L_{\rm syn} \gg L_{\rm ph}$ &
$\mathcal{E} \gg (\mathcal{E}\mathcal{B})^{1/2} \gg \mathcal{G} \gg \mathcal{B}$ \\
5 & & $k \gg 1$ & $x \gg 1$ & $L_{\rm ssc} \gg L_{\rm syn} \gg L_{\rm up} \gg L_{\rm ph}$ &
$\mathcal{E} \gg \mathcal{B} \gg \mathcal{G}$ \\
6 & & & $x \ll 1, ~kx \gg 1, ~ kx^2 \gg 1$ & $L_{\rm syn} \gg L_{\rm ssc} \gg L_{\rm ph} \gg L_{\rm up}$ &
$\mathcal{B} \gg \mathcal{E} \gg \mathcal{E}^2/\mathcal{B} \gg \mathcal{G}$ \\
7 & & & $x \ll 1, ~kx \gg 1, ~ kx^2 \ll 1$ & $L_{\rm syn} \gg L_{\rm ph} \gg L_{\rm ssc} \gg L_{\rm up}$ &
$\mathcal{B} \gg \mathcal{E} \gg \mathcal{G} \gg \mathcal{E}^2/\mathcal{B}$ \\
8 & & & $x \ll 1, ~kx \ll 1$ & $L_{\rm ph} \gg L_{\rm syn} \gg L_{\rm up} \gg L_{\rm ssc}$ &
$\mathcal{B} \gg \mathcal{G} \gg \mathcal{E}$ \\
\hline
9 & $\eta > \eta_* (\mathcal{G}>1)$ & 
$k' \ll 1, ~k'x \ll 1$ & $x \ll 1$ & $L_{\rm ph} \gg L_{\rm up} \gg L_{\rm syn} \gg L_{\rm ssc}$ &
$\mathcal{G} \gg {\rm max}(\mathcal{E},\mathcal{B})$ \\ 
\hline
\end{tabular}
\label{tab:cases}
\end{table*}

We require to estimate the cooling Lorentz factor $\gamma_c$ for specifying the lepton
energy distribution averaged over the dynamical timescale and the various emission luminosities.
Since the cooling rate for one lepton with Lorentz factor $\gamma$ is $P(\gamma) = (4/3) \sigma_T
c \gamma^2 (U'_B + U'_{\rm syn} + U'_{\rm ph})$, the cooling Lorentz factor can be estimated by
$\gamma_c m_e c^2 = P(\gamma_c) r_i/(2c\eta)$. This reduces to
\begin{equation}
\gamma_c \simeq \frac{3m_e \mathcal{R}}{4 m_p \tau_{l,i} (\eta/\eta_*)^{8/3}} \frac{1}{k(1+x)+1}.
\label{eq:gamma_c}
\end{equation}
If $k \gg 1$, we can take $k(1+x)+1 \approx k(1+x)$. Then Eq.~(\ref{eq:x}) reduces to 
$x = (\epsilon_e h/\epsilon_B)/(1+x)$, and we have
\begin{equation}
x \approx \left\lbrace\begin{array}{cl} 
    \sqrt{\epsilon_e h/\epsilon_B}, & (x \gg 1), \\
    \epsilon_e h/\epsilon_B,        & (x \ll 1). \\
\end{array}\right.  ~~~(k \gg 1)
\end{equation}
This is a general result for the case in which the radiation is dominated by synchrotron and 
SSC emissions, as developed by \citet{sari01}.
In the first case, $x \gg 1$, we have $kx^2 \gg kx \gg x \gg 1$, so that the order of the four emission
luminosities is found to be $L_{\rm ssc} \gg L_{\rm syn} \gg L_{\rm up} \gg L_{\rm ph}$. 
The condition $k \gg 1$ and $x \gg 1$ is equivalent to $\epsilon_d \epsilon_e h \gg \epsilon_d \epsilon_B
\gg (\eta/\eta_*)^{8/3}$. The second case, $x \ll 1$, should be divided into three sub-cases; 
($kx \gg 1$ and $kx^2 \gg 1$), ($kx \gg 1$ and $kx^2 \ll 1$), and ($kx \ll 1$). The orders of the luminosities 
in these cases are summarized in Table~\ref{tab:cases}, which are labeled as cases 5, 6, 7, and 8, 
respectively. In the Table we define $\mathcal{G} \equiv (\eta/\eta_*)^{8/3}$, 
$\mathcal{E} \equiv \epsilon_d \epsilon_e h$, and $\mathcal{B} \equiv \epsilon_d \epsilon_B$ for
clarification.

If $k \ll 1$, we can take $k(1+x)+1 \approx kx+1$ in Eq.~(\ref{eq:gamma_c}). This case should
be divided into further two cases, $kx \gg 1$ or $kx \ll 1$. We can take $k(1+x)+1 \approx kx$
in the former case and $k(1+x)+1 \approx 1$ in the latter case. In the former case, we have
\begin{equation}
x \approx \sqrt{\epsilon_e h/\epsilon_B} ~~~ (k \ll 1, kx \gg 1). 
\end{equation}
This is shown as case 4 in Table~\ref{tab:cases}. In the latter case, we have
\begin{equation}
x \approx \frac{\epsilon_d \epsilon_e h}{(\eta/\eta_*)^{8/3}} ~~~ (k \ll 1, kx \ll 1).
\label{eq:x_high}
\end{equation}
This case should be divided into three sub-cases; ($x \ll 1$), ($x \gg 1$ and $kx^2 \ll 1$),
and ($x \gg 1$ and $kx^2 \gg 1$). These are labeled as cases 1, 2, and 3, respectively in Table~\ref{tab:cases}.
We can calculate the cooling Lorentz factor $\gamma_c$ for each case by Eq.~(\ref{eq:gamma_c}). 

The ordering of the emission luminosities for various cases listed in Table~\ref{tab:cases} divides
the GRB emission models into two groups. Since the synchrotron and SSC emission components can have 
very broad spectra, so that a condition $L_{\rm ph} \gg {\rm max}(L_{\rm syn}, L_{\rm ssc})$ is necessary
for the photospheric emission to be dominant in the MeV energy range rather than the synchrotron or
SSC emission components. Thus cases 1, 2, and 8 are included in the photospheric 
emission models, while cases 3, 4, 5, 6, and 7 are included in the synchrotron-SSC 
emission models. In this paper we focus on the former cases, and in particular cases 
1 and 2, for which the UP emission component is dominant in the high-energy range.
We will see that cases 1 and 2 can be consistent with the spectra of some LAT GRBs in Section~\ref{sec:case}.
Here we show the description of the cooling Lorentz factor in cases 1 and 2,
\begin{eqnarray}
&& \gamma_c \simeq \frac{3 m_e \mathcal{R}}{4 m_p \tau_{l,i} (\eta/\eta_*)^{8/3}} \nonumber \\
&& \simeq 3\; L_{53}^{-1/3} \left(\frac{r_{a,7}}{\Gamma_a}\right)^{-2/3} \eta_{3}^{7/3}
\mathcal{R}_1^{2/3} t_{v,-2} \left(\frac{a_L}{5}\right)^{-2}.
\label{eq:gamma_c_high}
\end{eqnarray}
The UP luminosity is simply written as
\begin{equation}
L_{\rm up} \simeq L \epsilon_d \epsilon_e h.
\label{eq:Lup_high}
\end{equation}

\begin{figure*}
\begin{center}
\includegraphics{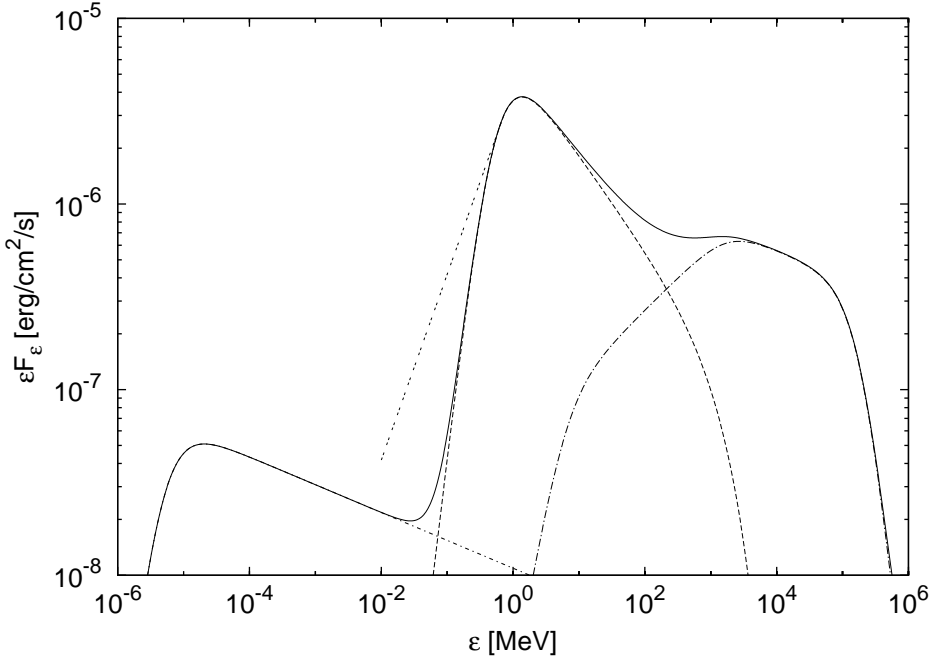}
\end{center}
\caption{
Example of the model spectrum of case 1 (see Table~\ref{tab:cases}). 
The spectrum consists of the photospheric emission (dashed line),
the UP emission (dot-long-dashed line), and the synchrotron emission (dot-short-dashed line). The SSC emission
is not shown since it is too dim. The parameters are $L_{53} = 3, \eta_3 = 3, r_{a,7}/\Gamma_a = 1,
\mathcal{R}_1 = 2, \beta_{\rm ph} = -2.5, t_{v,-2} = 1, p = 2.3, \epsilon_d \epsilon_e = 0.1,
\epsilon_d \epsilon_B = 0.03$, $a_L = 5$, and $z=2$. The characteristic quantities are 
$\eta_* \simeq 4.3 \times 10^3$,
$L_{\rm ph} \simeq L (\eta/\eta_*)^{8/3} \simeq 1 \times 10^{53}\;{\rm erg}\;{\rm s}^{-1}$,
$\varepsilon_{\rm ph} \simeq 4 k T_a (\eta/\eta_*)^{8/3} \simeq 4\;$MeV, 
$\varepsilon^{\rm ph}_{\gamma\gamma} \simeq 5\;$GeV, $L_{\rm up} \simeq 
L \epsilon_d \epsilon_e h \simeq 2 \times 10^{52}\;{\rm erg}\;{\rm s}^{-1}$,
$\varepsilon_{{\rm up},m} \simeq \varepsilon_{\rm ph} \gamma_m^2 \simeq 20\;$MeV, 
$\varepsilon_{{\rm up},c} \simeq \varepsilon_{\rm ph} \gamma_c^2 \simeq 5\;$GeV, 
$\varepsilon_{\rm up,KN} \simeq (\eta m_e c^2)^2/\varepsilon_{\rm ph} \simeq 600\;$GeV,
$L_{\rm syn} \simeq \epsilon_d \epsilon_B L_{\rm up} /(\eta/\eta_*)^{8/3} \simeq 1 \times 10^{51}\;{\rm erg}
\;{\rm s}^{-1}$, $\varepsilon_{{\rm syn},c} \simeq \varepsilon_{\rm syn}(\gamma_c) \simeq 80\;$eV,
and $\varepsilon_{{\rm syn},a} \simeq 60\;$eV.
The intra-source $e^{\pm}$ pair creation opacity is estimated to be less than unity.
We do not take into account the absorption by the $e^{\pm}$ pair creation through propagating in
the intergalactic medium.
The two-dot line describes a low-energy spectrum with an index $\alpha_{\rm ph} = -1.0$, typical
of observed GRBs, whose possible origins are discussed in section \ref{sec:summary}. 
}      
\label{fig:spec_ex}
\end{figure*}

Figure~\ref{fig:spec_ex} shows an example of a broadband spectrum of the emission from the GRB 
jet for case 1. The parameters are taken as $L_{53} = 3, \eta_3 = 3, r_{a,7}/\Gamma_a = 1,
\mathcal{R}_1 = 2, \beta_{\rm ph} = -2.5$, $t_{v,-2} = 1$, $p=2.3$, $\epsilon_d \epsilon_e = 0.1$,
$\epsilon_d \epsilon_B = 0.03$, and $a_L = 5$. The source redshift is set to be $z = 2$.
For these parameters, we have $(\eta/\eta_*)^{8/3} \simeq 0.4 > {\rm max}(\epsilon_d \epsilon_d h, 
\epsilon_d \epsilon_B)$, and $r_{\rm ph} \simeq 1 \times 10^{11}\;$cm and $r_i \simeq 
2 \times 10^{14}\;$cm, which satisfy our assumption $r_i \gg r_{\rm ph}$ (Eq.~\ref{eq:cond_high}).
We generate the spectra of the four emission components using the approximate forms 
of the individual components.
The photospheric emission is approximated as a smoothed broken power-law spectrum 
\begin{equation}
\varepsilon F_{\varepsilon}^{\rm ph} = A_{\rm ph} \left(\frac{\varepsilon}{\varepsilon_{\rm ph}}\right)^3
\left[1 + \left(\frac{\varepsilon}{\varepsilon_{\rm ph}}\right)^s \right]^{\frac{\beta_{\rm ph} - 1}{s}},
\end{equation}
where a constant $A_{\rm ph}$ is given for the peak of $\varepsilon F_{\varepsilon}^{\rm ph}$ to be equal to 
$L_{\rm ph}/(4\pi d_L^2)$ ($d_L$ is the luminosity distance) and we set $s=2$. 
The UP, synchrotron, and SSC components are approximated similarly as smoothed broken power-law spectra 
with the $\varepsilon F_{\varepsilon}$ peak values being equal to 
$L_{\rm up}/(4\pi d_L^2)$, $L_{\rm syn}/(4\pi d_L^2)$, and $L_{\rm ssc}/(4\pi d_L^2)$, respectively. 
The $\varepsilon F_{\varepsilon}$ spectral indices of the UP emission are given by
$2$ for $\varepsilon < \varepsilon_{{\rm up},l}$, $(3-q)/2$ for 
$\varepsilon_{{\rm up},l} < \varepsilon < \varepsilon_{{\rm up},h}$,
$(2-p)/2$ for $\varepsilon_{{\rm up},h} < \varepsilon < \varepsilon_{\rm up,KN}$,
and $-p-1$ for $\varepsilon > \varepsilon_{\rm up,KN}$, where
$\varepsilon_{{\rm up},l} \simeq \varepsilon_{\rm ph} \gamma_l^2$,
$\varepsilon_{{\rm up},h} \simeq \varepsilon_{\rm ph} \gamma_h^2$
($\gamma_l \equiv {\rm min}(\gamma_m, \gamma_c)$ and $\gamma_h \equiv {\rm max}(\gamma_m, \gamma_c)$),
$\varepsilon_{\rm up,KN} \simeq (\eta m_e c^2)^2/\varepsilon_{\rm ph}$, and $q = p ~(q = 2)$ for
the slow cooling case $\gamma_m < \gamma_c$ (for the fast cooling case $\gamma_c < \gamma_m$).
This is valid for the case of $\beta_{\rm ph} \leq -(2+p)/2$, which is satisfied for our case
$\beta \la -2.5$ and $2 < p < 3$.
The spectral indices of the synchrotron and SSC emission are found in \citet{sari01}.
Figure~\ref{fig:spec_ex} shows that the overall spectrum consists of a photospheric
component in the mid-range, and synchrotron and UP components in the low and high
energy ranges, respectively. We will use this approximate method to have a rough overall spectrum and
to show that the emission from the photosphere and internal shock can be consistent with the time-resolved
spectra of observed LAT GRBs, GRB 080916C, GRB 090510, and GRB 090902B in section~\ref{sec:case}.

The high-energy photospheric photons are absorbed by the $e^{\pm}$ pair creation at the photosphere.
Assuming that the photons are isotropic in the comoving frame of the jet,
the opacity for the photons with the comoving frame energy $\varepsilon'$ can be estimated by
$\tau^{\rm ph}_{\gamma\gamma}(\varepsilon') \simeq 0.1 \sigma_T \varepsilon'_{\rm ann} 
n'_{\rm ph}(r_{\rm ph}; \varepsilon'_{\rm ann}) r_{\rm ph}/(2\eta)$, 
where $\varepsilon'_{\rm ann} = m_e^2 c^4/\varepsilon'$, and $n'_{\rm ph}(r_{\rm ph}; \varepsilon')
\simeq L_{\rm ph}(\varepsilon'/\varepsilon'_{\rm ph})^{\beta_{\rm ph}}/(4\pi r_{\rm ph}^2 c \eta^2
{\varepsilon'}^2_{\rm ph})$ is the comoving photon density per unit energy around 
$\varepsilon'$ at the photosphere \citep[cf.,][]{lithwick01}. 
Here we only consider the high-energy part of the photospheric emission (with the index $\beta_{\rm ph}$) 
as target photons, since the photon density $\sim \varepsilon' n'_{\rm ph}(r_{\rm ph}; \varepsilon')$ peaks 
at $\varepsilon'_{\rm ph}$. If a calculated main target photon (comoving) energy for the pair creation
break energy is smaller than $\varepsilon'_{\rm ph}$, the region is optically thin for the pair creation.
The pair creation break energy, defined by $\tau^{\rm ph}_{\gamma\gamma}(\varepsilon') = 1$, is given by
\begin{eqnarray}
\varepsilon^{\rm ph}_{\gamma\gamma} &\simeq& \left[0.1\; \frac{m_p}{m_e} \left(\frac{\eta}{\eta_*}\right)^{8/3}
\mathcal{R}^{-1} \eta^{2\beta_{\rm ph} +3} \right. \nonumber \\
&& \left. \left(\frac{\varepsilon_{\rm ph}}{m_e c^2}\right)^{-2-\beta_{\rm ph}} 
\right]^{\frac{1}{1+\beta_{\rm ph}}} \times m_e c^2 \nonumber \\
&\simeq& 3\; \left[(9 \times 10^3)^{\frac{3+\beta_{\rm ph}}{2}} \left(\frac{\eta}{\eta_*}\right)^{8/3}
\mathcal{R}_1^{-1} \eta_3^{2\beta_{\rm ph}+3} \right. \nonumber \\
&& \left. \left(\frac{\varepsilon_{\rm ph}}{1\;{\rm MeV}}
\right)^{-2-\beta_{\rm ph}} \right]^{\frac{1}{1+\beta_{\rm ph}}} \;{\rm GeV}.
\label{eq:gammagamma_ph}
\end{eqnarray}
For the spectral model of the photospheric emission, we assume a spectral cutoff at this energy.

The $e^{\pm}$ pair creation opacity for the high-energy UP photons at the internal shock
region can be estimated similarly. The main target photons for the high-energy UP photons may be either 
the photospheric emission incident into the internal shock region or the UP emission itself.
Since the incident photospheric emission is highly anisotropic at the internal shock region and the 
collision angle of the two photons is typically very small, so that the pair creation opacity 
will be significantly reduced from that with the isotropic assumption 
\citep{zou10} \citep[see also][]{granot08,ackermann090926A}.
Yet, for simplicity, we estimate the pair creation opacity by the interaction with the photospheric 
emission with the isotropic assumption, which provides a possible largest opacity. On the other hand, the 
target UP photons are much less anisotropic. We derive a minimum possible value of the 
break energy, which can be obtained as the lower one of the break energies caused by pair 
creations with the photospheric and the UP photons calculated under the isotropic assumption.
The photon densities per unit energy in the relevant energy range are given by 
$n'_{\rm ph}(r_i; \varepsilon') \simeq L_{\rm ph} (\varepsilon'/\varepsilon'_{\rm ph})^{\beta_{\rm ph}} / 
(4\pi r_i^2 c \eta^2 {\varepsilon'}^2_{\rm ph})$ and $n'_{\rm up}(r_i; \varepsilon') \simeq L_{\rm up}
(\varepsilon'/\varepsilon'_{\rm up})^{-(1+q)/2} / (4\pi r_i^2 c \eta^2 {\varepsilon'}^2_{{\rm up},h})$
for the photospheric and UP emission, respectively. Thus the possible minimum break energy is estimated 
by $\varepsilon_{\gamma\gamma}^{\rm min} \simeq {\rm min}(\varepsilon_{\gamma\gamma,{\rm ph}}^{\rm min},
~\varepsilon_{\gamma\gamma,{\rm up}}^{\rm min})$ and
\begin{eqnarray}
\varepsilon_{\gamma\gamma,{\rm ph}}^{\rm min} &\simeq& 
\left[\frac{3}{40} \gamma_c^{-1} \eta^{2\beta_{\rm ph}+3} 
\left(\frac{\varepsilon_{\rm ph}}{m_e c^2} \right)^{-2-\beta_{\rm ph}} \right]^{\frac{1}{1+\beta_{\rm ph}}}
\times m_e c^2 \nonumber \\
&\simeq& 1 \times 10^2\; \left[4^{\frac{3+\beta_{\rm ph}}{2}} \left(\frac{\gamma_{c}}{10} \right)^{-1} 
\eta_3^{2\beta_{\rm ph}+3} \right. \nonumber \\
&& \left. \left(\frac{\varepsilon_{\rm ph}}{1\;{\rm MeV}}\right)^{-2-\beta_{\rm ph}}
\right]^{\frac{1}{1+\beta_{\rm ph}}} \; {\rm GeV}, \nonumber \\
\varepsilon_{\gamma\gamma,{\rm up}}^{\rm min} &\simeq& 
\left[\frac{3}{40} \gamma_c^{-1} x \eta^{2-q} 
\left(\frac{\varepsilon_{{\rm up},h}}{m_e c^2} \right)^{\frac{q-3}{2}} \right]^{\frac{2}{1-q}}
\times m_e c^2 \nonumber \\
&\simeq& 2 \times 10^1\; \left[(4 \times 10^{-3})^{2-q} \left(\frac{\gamma_c}{10} \right)^{-1}
x \eta_3^{2-q} \right. \nonumber \\
&& \left. \left(\frac{\varepsilon_{{\rm up},h}}{1\;{\rm GeV}} \right)^{\frac{q-3}{2}} 
\right]^{\frac{2}{1-q}}\; {\rm TeV},
\label{eq:gammagamma}
\end{eqnarray}
where $x$ is given by Eq.~(\ref{eq:x_high}). The opacity of the interaction with the photospheric
emission (the UP emission) for all the high-energy photons is less than unity if 
$\varepsilon_{\gamma\gamma,{\rm ph}}^{\rm min} > (\eta^2 m_e c^2)^2/\varepsilon_{\rm ph}$
($\varepsilon_{\gamma\gamma,{\rm up}}^{\rm min} > (\eta^2 m_e c^2)^2/\varepsilon_{{\rm up},l}$).

The energy $\varepsilon_{{\rm syn}, a}$ below which the synchrotron self-absorption effect
is significant can be estimated by equating synchrotron flux to the blackbody flux of 
the characteristic electrons in the shocked region, $F_{\varepsilon_{{\rm syn}},a} =
[(1+z)^3/d_L^2] 2\pi m_e \gamma_{ch} (\varepsilon_{{\rm syn},a}/h)^2(r_i^2/\eta)$
\citep[e.g.,][]{meszaros97}. The characteristic Lorentz factor $\gamma_{ch}$ of leptons is
given by $\gamma_a$ whose synchrotron energy is $\varepsilon_{{\rm syn},a}$ in the case of
$\gamma_a > \gamma_l$ and otherwise by $\gamma_l$.
We show a formula of the ratio of $\varepsilon_{{\rm syn},a}$ to $\varepsilon_{{\rm syn},h}
= \varepsilon_{\rm syn}(\gamma_h)$, by which we can calculate
$\varepsilon_{{\rm syn},a}$ in a typical case $\gamma_a > \gamma_l$,
\begin{equation}
\frac{\varepsilon_{{\rm syn},a}}{\varepsilon_{{\rm syn},h}} \simeq
\left[\frac{8\pi \sqrt{3}}{9} \frac{e \tau_{l,i}}{\sigma_T B'} \gamma_l^{q-1} \gamma_h^{-q-4} \right]^w,
\label{eq:ssa}
\end{equation}
where $w = 2/(q+4)$ for $\gamma_l < \gamma_a < \gamma_h$ and $w = 2/(p+5)$ for $\gamma_a > \gamma_h$.
The value of $\varepsilon_{{\rm syn},a}$ is important for estimating the observed synchrotron
flux at the optical frequency (see the above example shown in Figure~\ref{fig:spec_ex}).

\subsection{Case of $\eta > \eta_*$}

In this case the kinetic luminosity that can be dissipated in the internal shock is 
$L_k \approx L(\eta/\eta_*)^{-1}$. The internal shock radius is estimated to be 
$r_i \simeq 2 c t_v \eta_*^2 a_L^{-2}$, and our assumption $r_i \gg r_f$ reduces to
\begin{equation}
\frac{r_a}{\Gamma_a} \ll 2 c t_v \eta_* a_L^{-2} 
\simeq 6 \times 10^{11}\;t_{v,-2} \eta_{*,3} a_L^{-2} \;{\rm cm}.
\label{eq:cond_low}
\end{equation}
The minimum injection Lorentz factor of the leptons is 
given by Eq.(\ref{eq:gamma_m}). The magnetic field energy density is $U'_B = L\epsilon_d \epsilon_B/
(4\pi r_i^2 c \eta \eta_*)$, and then the characteristic synchrotron energy of leptons with Lorentz
factor $\gamma$ is given by
\begin{eqnarray}
\varepsilon_{\rm syn}(\gamma) &\simeq& \frac{3heB'}{4\pi m_e c} \gamma^2 \eta_* \nonumber \\ 
&\simeq& 5 \times 10^{-1}\;
\gamma^2\; L_{53}^{1/8} \left(\frac{r_{a,7}}{\Gamma_a}\right)^{3/8} \eta_{3}^{-1/2} \mathcal{R}_1^{-3/8} \nonumber \\
&& \times\; t_{v,-2}^{-1} a_L^2 \left(\frac{\epsilon_d \epsilon_B}{0.1}\right)^{1/2}\; {\rm eV}.
\end{eqnarray}

The photospheric, UP, synchrotron, and SSC luminosities are estimated to be 
$L_{\rm ph} \approx L$, $L_{\rm up} \simeq x L$, $L_{\rm syn} \simeq k' x L$, and $L_{\rm ssc}
\simeq k' x^2 L$, where $x$ is calculated by Eq.(\ref{eq:x}), and we have defined
\begin{equation}
k' \equiv \frac{\epsilon_d \epsilon_B}{\eta/\eta_*}.
\end{equation}
The cooling Lorentz factor of leptons is
\begin{equation}
\gamma_c \simeq \frac{3 m_e \mathcal{R}}{4 m_p \tau_{l,i} (\eta/\eta_*)} \frac{1}{k'(1+x)+1},
\end{equation}
where the optical depth at $r_i$ can be written as $\tau_{l,i} = (r_{\rm ph}/r_i)(\eta/\eta_*)^{-2/3}$.
The equations for the luminosities and $\gamma_c$ are the same as those for the case $\eta < \eta_*$
by replacing $(\eta/\eta_*)^{8/3}$ by $\eta/\eta_*$ and $k$ by $k'$, so that the same argument for 
dividing cases for the order of the four luminosities can be made. In this case, however, 
$\eta > \eta_*$ leads to $k' < 1$ and $\epsilon_d \epsilon_e h/(\eta/\eta_*) < 1$. Then we only
have the case of $k' \ll 1$ and $k'x \ll 1$, which we include as case 9 in Table~\ref{tab:cases}.
In this case the MeV and high-energy emissions may be dominated by the photospheric and UP
components, respectively, similar to case 1.
(The case of $k' \ll 1$ and $k' x \gg 1$ leads to $x \approx \sqrt{\epsilon_e h /\epsilon_B}~(\gg 1)$, 
which is not consistent with the condition $k' x^2 = \epsilon_d \epsilon_e h / (\eta/\eta_*) < 1$.)
The cooling Lorentz factor is calculated by
\begin{equation}
\gamma_c \simeq \frac{3 m_e \mathcal{R}}{4 m_p \tau_{l,i} (\eta/\eta_*)}
\simeq 7\times 10^2 \; L_{53}^{1/4} \left(\frac{r_{a,7}}{\Gamma_a}\right)^{-5/4} \mathcal{R}_1^{5/4} 
t_{v,-2} a_L^{-2}.
\label{eq:gamma_c_low}
\end{equation}
The parameter $x$ and the UP luminosity are given by
\begin{eqnarray}
x &\approx& \frac{\epsilon_d \epsilon_e h}{\eta/\eta_*}, 
\label{eq:x_low} \\
L_{\rm up} &\simeq& L \left(\frac{\eta}{\eta_*}\right)^{-1} \epsilon_d \epsilon_e h.
\label{eq:Lup_low}
\end{eqnarray}

The pair creation break energies of the photospheric and the UP emission can be estimated by 
Eq.~(\ref{eq:gammagamma_ph}) with replacing $(\eta/\eta_*)^{8/3}$ by $(\eta/\eta_*)^{-(8/3)(1+\beta_{\rm ph})}$
and Eq.~(\ref{eq:gammagamma}) with replacing $\eta$ by $\eta_*$ and setting $x$ to be
Eq.~(\ref{eq:x_low}), respectively. The synchrotron self-absorption energy can be estimated by
the same equation as Eq.~(\ref{eq:ssa}).

\section{Case studies}
\label{sec:case}

The previous sections have provided
a general formulation of the emission from the photosphere and internal 
shock of the GRB jet. Here we focus on the cases in which the emission in the MeV 
energy range and in the high-energy range are dominated by the photospheric and 
the UP emission components, respectively,
instead of the synchrotron or SSC emission components
(i.e., cases 1, 2, and 9 in Table~\ref{tab:cases}). We then show that such cases can be
consistent with the observed time-resolved spectra of three of the best 
observed {\it Fermi}/LAT GRBs, especially in the energy range at and above the MeV 
spectral peaks.  The detailed analysis results of the observed spectra of brightest 
LAT GRBs, GRB 080916C, GRB 090902B and GRB 090510 have been published 
\citep{abdo080916C,abdo090902B,ackermann090510}, 
and it is to these data that we apply our formulation. We assume that essentially
the parameters $W$ and $c t_v$ are in the efficient scattering regime 
(see section~\ref{sec:temporal}), unless otherwise stated.

\subsection{GRB 080916C}

\begin{figure*}
\begin{center}
\includegraphics{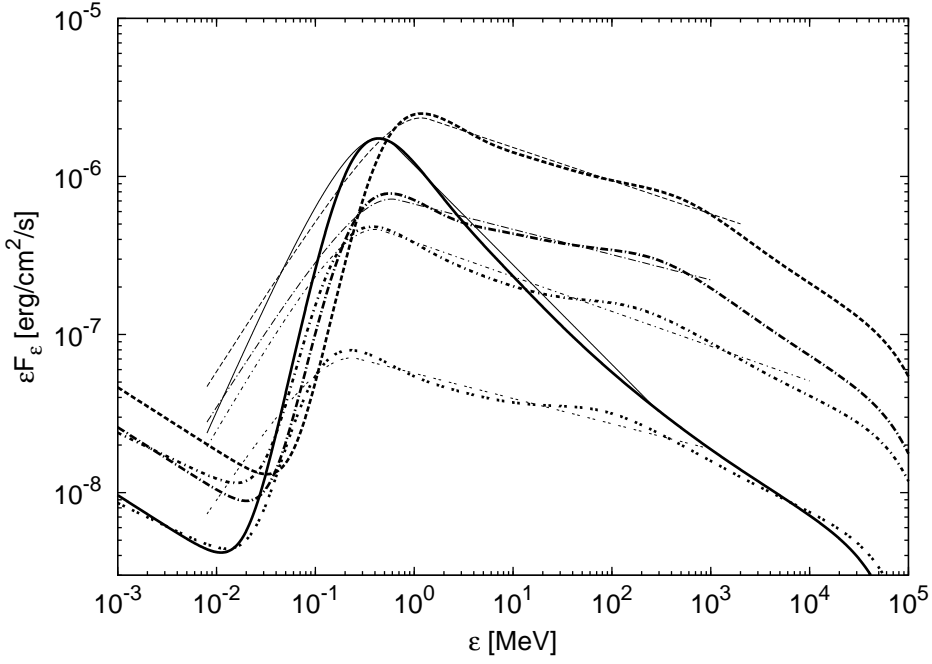}
\end{center}
\caption{
Model fits of the time-resolved spectra of GRB 080916C. The thin lines are the observed data,
and the thick lines are the model fits (solid lines are for $0.0-3.6\;$s, dashed
lines are for $3.6-7.7\;$s, dot-long-dashed lines are for $7.7-15.8\;$s, dot-short-dashed
lines are for $15.8-54.7\;$s, and the two-dot lines are for $54.7-100.8\;$s). Adopted parameter values
are listed in Table~\ref{tab:parameters}. The spectral portions
at and above the peak energies are well described by the model. The observed spectral parts below
the peak energies are significantly softer than the model. This issue is discussed in 
Section~\ref{sec:summary}.
}      
\label{fig:spec_080916C}
\end{figure*}

This burst is a long GRB that occurred at a redshift $z \simeq 4.35$ (corresponding to
$d_L \simeq 1.2\times10^{29}\;$cm).  The spectral analysis shows that all the spectra of the 
five time-bins can be fitted by Band functions \citep{abdo080916C}. These time-binned 
spectra are shown by the thin lines in Figure~\ref{fig:spec_080916C}. 
The high-energy spectral index of the second time-bin $3.6-7.7\;$s, $\beta \simeq -2.2$, is 
significantly larger than that of the first time-bin $0.0-3.6\;$s, $\beta \simeq -2.6$. 
This spectral hardening corresponds to the observed delay of the onset of the LAT emission 
with respect to that of the GBM emission. The delay timescale in the cosmological rest frame is 
$\sim 5/(1+z)\;{\rm s} \sim 1\;$s. The high-energy spectral index seems stable after the second time-bin. 
We will see that the first time-bin spectrum is consistent with having only a photospheric 
component, while the subsequent time-bin spectra can be modeled as a photospheric plus a UP component,
which mimic a smooth Band function. 

We first find the parameter values appropriate for the {\it second} time-bin spectrum. 
In our adiabatic jet model, we have a constraint $\beta_{\rm ph} \la -2.5$ (see Section~\ref{sec:ph}).
Thus the observed hard high-energy spectrum of this time-bin with $\beta \simeq -2.2$ has to be made by 
a combination of the photospheric and UP emission. 
The spectrum can be fitted in this way as shown by the thick dashed line in Figure~\ref{fig:spec_080916C}. 
The following conditions are required for producing such a spectrum: 
$L_{\rm ph} \sim 4.5 \times 10^{53}\;{\rm erg}\;{\rm s}^{-1}$, $\varepsilon_{\rm ph} \sim 6.3\;$MeV,
$\beta_{\rm ph} \simeq -2.5$, $L_{\rm up} \sim 1.1 \times 10^{53}\;{\rm erg}\;{\rm s}^{-1}$,
$\varepsilon_{{\rm up},c} \sim 2.5\;$GeV, $\varepsilon_{{\rm up},m} \la 42\;$MeV (which correspond to
$\gamma_c \sim 20$ and $\gamma_m \la 2.6$), and $p \simeq 2.8$. Significant deviations from these
conditions lead to a bumpy spectrum and/or deviate from the observed data with errors taken into account
(see below for other conditions). The conditions indicate the slow-cooling
case of the electron distribution, i.e., $\gamma_m < \gamma_c$. The fast-cooling case is not favored
since it has a harder spectrum at $\varepsilon_{{\rm up},c} < \varepsilon < \varepsilon_{{\rm up},m}$ 
which makes a dip in the spectrum. 
The ratio of the two luminosities is $L_{\rm up}/L_{\rm ph} = x \sim 0.2$, which 
indicates $\eta < \eta_*$ for a reasonable value of $\epsilon_d \epsilon_e \la 0.1$ for the internal
shock (see Equations~\ref{eq:x_high} and \ref{eq:x_low}). Thus this spectrum corresponds to case 1
(see Table~\ref{tab:cases}). Equations for case 1, (\ref{eq:ph_high_baryon}), (\ref{eq:Lup_high}), 
(\ref{eq:gamma_m}), and (\ref{eq:gamma_c_high}), translate the above conditions into five constraints
on the model parameters,
\begin{eqnarray}
&& L_{53} \left(\frac{\eta}{\eta_*}\right)^{8/3} \sim 4.5 \left(\frac{L_{\rm ph}}{4.5 \times 10^{53}\;
{\rm erg}\;{\rm s}^{-1}}\right), \label{eq:const1} \\
&& L_{53} \left(\frac{\epsilon_d \epsilon_e}{0.1}\right) \sim 11\;h^{-1} \left(\frac{L_{\rm up}}{1.1 \times
10^{53}\;{\rm erg}\;{\rm s}^{-1}}\right), \label {eq:const2} \\
&& L_{53}^{1/4} \left(\frac{r_{a,7}}{\Gamma_a}\right)^{-1/2} \left(\frac{\eta}{\eta_*}\right)^{8/3}
\sim 0.75 \left(\frac{\varepsilon_{\rm ph}}{6.3\;{\rm MeV}}\right), \label{eq:const3} \\
&& \mathcal{R}^{-1}_1 \left(\frac{\epsilon_d \epsilon_e}{0.1}\right) \sim 0.32 
\left(\frac{\gamma_m}{2.6}\right) \left(\frac{f(p)}{1.9}\right)^{-1}, \label{eq:const4} \\
&& L_{53}^{-1/3} \left(\frac{r_{a,7}}{\Gamma_a}\right)^{-2/3} \eta_3^{7/3} \mathcal{R}_1^{2/3}
t_{v,-2} \left(\frac{a_L}{5}\right)^{-2} \nonumber \\
&& ~~~~~~~\sim 7.9 \left(\frac{\gamma_c}{20}\right). \label{eq:const5}
\end{eqnarray}
We can constrain the parameters $L$, $\eta$, $r_a/\Gamma_a$, $\mathcal{R}$, and $t_v$ through these 
equations, by choosing reasonable values $\epsilon_d \epsilon_e \sim 0.1$ and $a_L \sim 5$ for 
internal shock. The parameters $\beta_{\rm ph}$ and $p$ have already been constrained directly by the
model fit. The remaining parameter $\epsilon_d \epsilon_B$ does not affect the photospheric 
and UP spectra and it is constrained by the synchrotron emission contribution in the low 
energy range (see below). The above first three equations (\ref{eq:const1}, \ref{eq:const2}, and
\ref{eq:const3}) provide $L_{53} \sim 11 h^{-1}$, $(\eta/\eta_*)^{8/3} \sim 0.41 h$, and
$r_{a,7}/\Gamma_a \sim 0.98 h^{3/2}$. We have some allowed ranges of parameters depending on the 
value of $\gamma_m$. For $\gamma_m \sim 2.6$, we have $h \sim 0.98$, and Equations (\ref{eq:const4})
and (\ref{eq:const5}) lead to $\mathcal{R} \sim 3$, $\eta_3 \sim 4.9$, and $t_{v,-2} \sim 0.2$, where 
Equation (\ref{eq:eta_*}) is used. If we adopt a smaller value, e.g., $\gamma_m \sim 1$, we have 
$h \sim 0.45$, leading to $\mathcal{R} \sim 8$, $\eta_3 \sim 7.4$, and $t_{v,-2} \sim 0.02$. 
The synchrotron emission fluxes, which are dominant below $10\;$keV,
should be smaller than the observed fluxes around $10\;$keV, which provide upper bounds
$\epsilon_d \epsilon_B \la 0.1$ for both cases. 
A larger $h$ case has a smaller luminosity budget $L$.
Comparisons of the prompt emission with the afterglow in GRBs generally imply high radiation
efficiencies of the prompt emission, $\epsilon_\gamma \ga 0.5$ \citep[e.g.,][]{panaitescu02}. 
Thus we favor the case of the 
allowed smallest luminosity, i.e., the case of $\gamma_m \sim 2.6$, which we list the constrained 
parameter values in Table~\ref{tab:parameters}.

\begin{table*}
\centering
\begin{minipage}{170mm}
\caption{Best fit model parameters for GRB 080916C, GRB 090902B, and GRB 090510}
\begin{tabular}{cccccccccccc}
\hline
GRB 080916C\footnote{
These values are example sets of parameter values for producing the fitting spectra, for which 
$L$ is the allowed smallest values for each time-bin. See texts for detailed explanations how to
specify the other values.
Two sets of the parameter values are shown for the time-bins of $0.0-3.6\;$s of GRB 080916C, $0.5-0.6\;$s
and $0.8-0.9\;$s of GRB 090510. The parameter $a_L$ is taken as $5$ for all the time-bins except
that we take $a_L = 2$ for the second set for the $0.0-3.6\;$s time-bin of GRB 080916C and
for the second set for the $0.5-0.6\;$s time-bin of GRB 090510.
} 
& Time-bin & $L_{53}$ & $r_{a,7}/\Gamma_a$ & $\eta_3$ & $\mathcal{R}_1$ & $\beta_{\rm ph}$ &
$t_{v,-2}$ & $p$ & $\epsilon_d \epsilon_e$ & $\epsilon_d \epsilon_B$ & $\eta/\eta_*$ \\
\hline 
 & $0.0-3.6\;$s & $3$ & $20$ & $3.2$ & $2$ & $-3.0$ & $<1.0$ & $2.8$ & $0.1$ & $<1$ & $1.6$ \\
 &              & $3$ & $20$ & $3.1$ & $2$ & $-2.7$ & $0.1$ & $2.8$ & $0.01$ & $<1$ & $1.6$ \\ 
 & $3.6-7.7\;$s & $10$ & $0.9$ & $4.9$ & $3$ & $-2.5$ & $0.2$ & $2.8$ & $0.1$ & $\la 0.1$ & $0.71$ \\
 & $7.7-15.8\;$s & $5$ & $1$ & $3.2$ & $3$ & $-2.5$ & $0.6$ & $2.8$ & $0.1$ & $\la 0.1$ & $0.61$ \\
 & $15.8-54.7\;$s & $3$ & $1$ & $2.5$ & $3$ & $-2.5$ & $0.7$ & $2.6$ & $0.1$ & $\la 0.1$ & $0.52$ \\
 & $54.7-100.8\;$s & $0.7$ & $2$ & $1.6$ & $3$ & $-2.5$ & $1.5$ & $2.6$ & $0.1$ & $\la 0.2$ & $0.55$ \\
\hline
\hline
GRB 090902B$^{a}$ & Time-bin & $L_{53}$ & $r_{a,7}/\Gamma_a$ & $\eta_3$ & $\mathcal{R}_1$ & $\beta_{\rm ph}$ &
$t_{v,-2}$ & $p$ & $\epsilon_d \epsilon_e$ & $\epsilon_d \epsilon_B$ & $\eta/\eta_*$ \\
\hline 
 & $4.6-9.6\;$s & $16$ & $2$ & $3.5$ & $3$ & $-4.0$ & $10$ & $2.9$ & $0.1$ & $0.1$ & $0.52$ \\
\hline
\hline
GRB 090510$^{a}$ & Time-bin & $L_{53}$ & $r_{a,7}/\Gamma_a$ & $\eta_3$ & $\mathcal{R}_1$ & $\beta_{\rm ph}$ &
$t_{v,-2}$ & $p$ & $\epsilon_d \epsilon_e$ & $\epsilon_d \epsilon_B$ & $\eta/\eta_*$ \\
\hline 
 & $0.5-0.6\;$s & $1$ & $3$ & $73$ & $2$ & $-4.8$ & $0.3$ & $2.3$ & $0.1$ & $<1$ & $26$ \\
 &              & $1$ & $3$ & $5.4$ & $2$ & $-4.8$ & $0.1$ & $2.3$ & $5 \times 10^{-3}$ & $<1$ & $1.9$ \\
 & $0.6-0.8\;$s & $4$ & $0.3$ & $4.7$ & $2$ & $-3.0$ & $0.2$ & $2.3$ & $0.1$ & $\la 7\times10^{-3}$ & $0.72$ \\
 & $0.8-0.9\;$s & $20$ & $2 \times 10^{-3}$ & $3.8$ & $0.5$ & $-2.8$ & $0.2$ & $2.05$ & $0.15$ & $\la 3\times10^{-6}$ & $0.18$\\
 &              & $10$ & $0.01$ & $1.6$ & $0.6$ & $-2.5$ & $7$ & $2.15$ & $0.15$ & $\la 4\times10^{-6}$ & $0.12$\\
\hline
\end{tabular}
\label{tab:parameters}
\end{minipage}
\end{table*}

Next we consider the {\it first} time-bin spectrum. If the internal shock for this 
component produces an electron energy distribution with $p \simeq 2.8$ similar to the second 
time-bin, the UP emission would have a hard photon index $-(p+2)/2 \simeq -2.4$ in the GeV energy range.
Thus the UP emission has to be very dim compared with the photopheric emission, and the overall spectrum
would consist mainly of the photospheric emission. The luminosity ratio should be 
$L_{\rm up}/L_{\rm ph} = x \la 0.06$. The internal shock may have $\epsilon_d \epsilon_e \sim 0.1$ similar
to the second time-bin, so that we require $\eta \ga \eta_*$ (see Equations~\ref{eq:x_high} and 
\ref{eq:x_low}). Thus this spectrum corresponds to case 9 (see Table~\ref{tab:cases}).
As shown by the thick solid line in Figure~\ref{fig:spec_080916C}, we can fit the spectrum mainly 
by the photospheric emission whose luminosity, peak energy, and high-energy spectral index are set 
to be $L_{\rm ph} \sim 3.1 \times 10^{53}\;{\rm erg}\;{\rm s}^{-1}$, $\varepsilon_{\rm ph} \sim 2.3\;$MeV,
and $\beta_{\rm ph} \sim -3.0$, respectively. Equations (\ref{eq:ph_low_baryon}) determine the parameters 
$L$ and $r_a/\Gamma_a$ by $L_{\rm ph}$ and $\varepsilon_{\rm ph}$, $L_{53} \simeq L_{{\rm ph},53} \sim 3$
and $r_{a,7}/\Gamma_a \sim 20$. The allowed ranges of $\eta$, $\mathcal{R}$, and $t_v$ are broad since 
the UP spectrum is not tightly constrained.
Equation~(\ref{eq:Lup_low}) means that larger $\eta$ decreases the UP luminosity. The model fit shown in 
Figure~\ref{fig:spec_080916C} is a case with the maximum contribution of the UP emission, i.e., a case of 
the minimum $\eta$ for the fixed values $\epsilon_d \epsilon_e \sim 0.1$ and $p \simeq 2.8$.
This model fit requires the luminosity and characteristic energies of the UP emission 
to be $L_{\rm up} \sim 2.0 \times 10^{52}\;{\rm erg}\;{\rm s}^{-1}$, 
$\varepsilon_{{\rm up},h} \sim 27\;$MeV, and $\varepsilon_{{\rm up},l} \la 10\;$MeV, respectively
(corresponding to $\gamma_h \sim 3.4$ and $\gamma_l \la 2.1$). Equations~(\ref{eq:gamma_m}) and
(\ref{eq:gamma_c_low}) indicate that we can have similar values of the model parameters $\mathcal{R}$
and $t_v$ as for the second time-bin, i.e., we have $\gamma_m \sim 3.4$ and $\gamma_c \la 2.1$ 
for $\mathcal{R}_1 \sim 2.4$ and $t_{v,-2} \la 1.0$. 
Since $\gamma_c < \gamma_m$, we have $h=1$, so that Equation~(\ref{eq:Lup_low}) provides $\eta_3 \sim 3.2$.

Alternatively, we may consider a case of $\epsilon_d \epsilon_e \ll 0.1$ for the first time-bin.
For the case of $\eta > \eta_*$, the final Lorentz factors of the shells are given by $\eta_*$,
which may not be much different, i.e., $a_L \la 2$, since it depends weakly on $L$, $r_a/\Gamma_a$, 
and $\mathcal{R}$. Then the internal shocks may only cause a weak dissipation, with 
$\epsilon_d \epsilon_e \ll 0.1$. The observed first time-bin spectrum can be fitted 
by a photospheric emission only, with $\beta_{\rm ph} \simeq -2.7$ in this case.
The model parameters $L$ and $r_a/\Gamma_a$ are constrained to be the same as the above case
$\epsilon_d \epsilon_e \sim 0.1$. For $\epsilon_d \epsilon_e \sim 0.01$ as an example, we can
find a set of parameter values $\eta$, $\mathcal{R}$, and $t_v$ similar to those for the 
second time-bin, which are shown in Table~\ref{tab:parameters}.

The spectral shapes of the other three time-bins are very similar to that of the second time-bin.
They can also be produced by a combination of the photospheric and UP components in our model,
where the luminosity ratios of the two components are required to be relatively large, i.e., $x \sim 0.3$,
which indicate case 1, similar to the second time-bin. We can constrain the model parameter values for 
the three time-bin spectra in a similar way to the second time-bin spectrum.
The parameters constrained for the case of the allowed smallest luminosity budgets are summarized in 
Table~\ref{tab:parameters}. The rightmost column represents the value of $\eta/\eta_*$ calculated for 
the adopted parameter values.

\begin{figure*}
\begin{tabular}{cc}
\begin{minipage}{0.5\hsize}
\begin{center}
\includegraphics[scale=0.65]{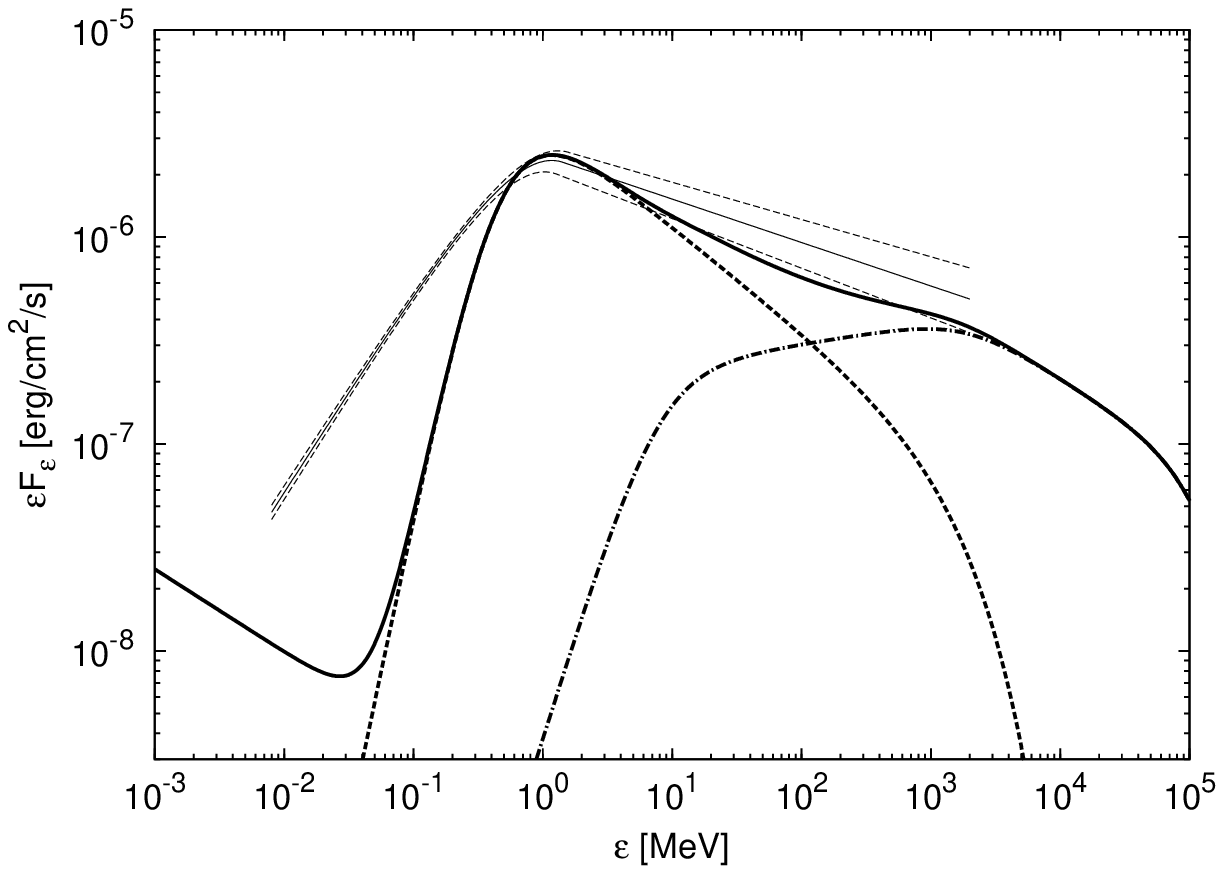}
\end{center}
\end{minipage}
\begin{minipage}{0.5\hsize}
\begin{center}
\includegraphics[scale=0.65]{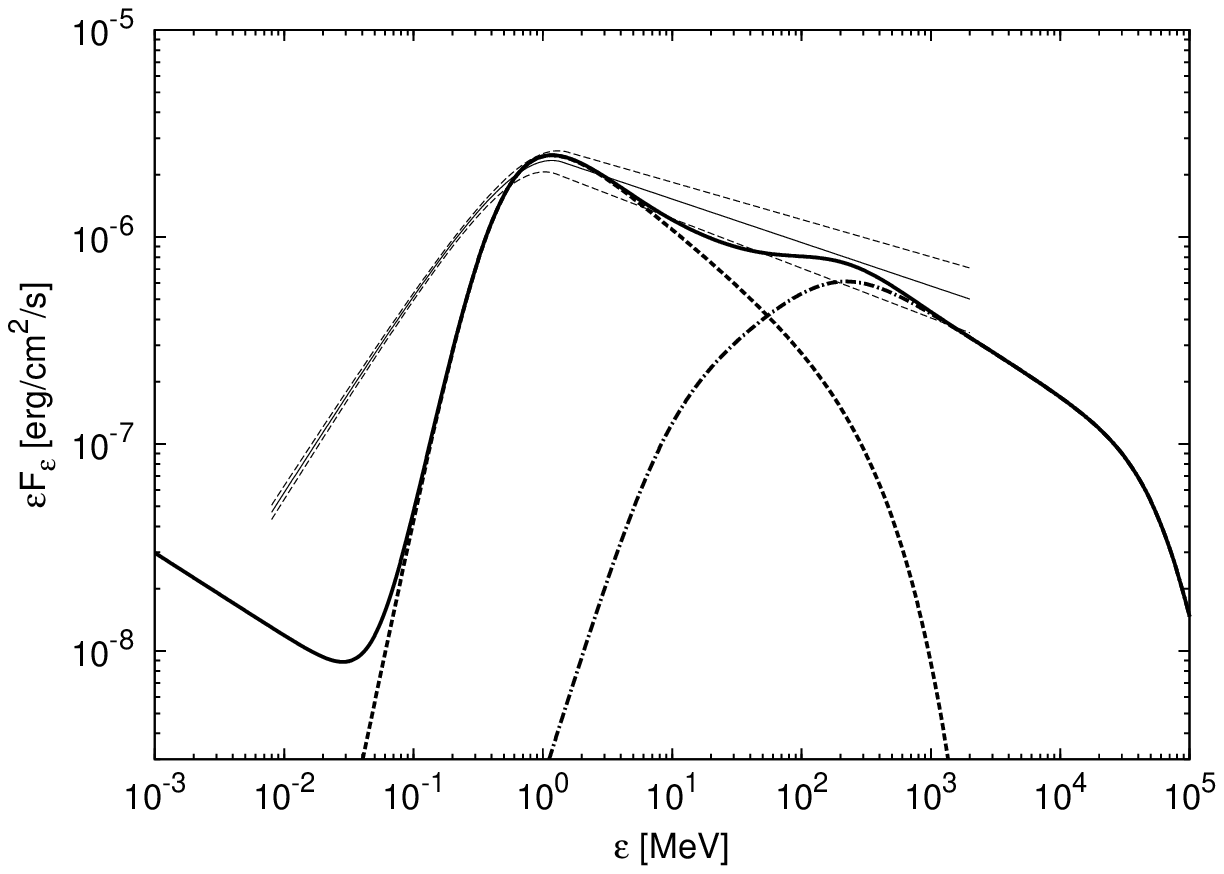}
\end{center}
\end{minipage}
\end{tabular}
\caption{
Model spectra for the second time-bin spectrum of GRB 080916C, for different values of 
$t_v$ and $\mathcal{R}$ and fixed values of $L$, $r_a/\Gamma_a$, $(\eta/\eta_*)^{8/3}$, 
$\beta_{\rm ph}$, $p$, $\epsilon_d \epsilon_e$ and $\epsilon_d \epsilon_B$ as shown in 
Table~\ref{tab:parameters}.
Left: The thick solid line is the model spectrum using the best fit parameter values shown
in Table~\ref{tab:parameters} except for taking $t'_v = 2 t_v$, which is compared with the thin lines
of the Band function fitting the observed data, with the dashed lines showing the $1\sigma$ errors
\citep{abdo080916C}.
Right: The thick solid line is the model spectrum using the best fit parameter values shown
in Table~\ref{tab:parameters} except for taking $\mathcal{R}' = \mathcal{R}/5$, which is 
compared with the same observed data as the Left panel. In both panels the thick dashed lines
and the thick dot-dashed lines represent the photospheric and UP components, respectively.
}      
\label{fig:spec_dp}
\end{figure*} 

The $e^{\pm}$ pair-creation 
break energies under the isotropic photon field assumption is given by
$\varepsilon_{\gamma\gamma,{\rm min}}^{\rm ph}/(1+z) \sim 10\;$GeV for the first time-bin and 
$\varepsilon_{\gamma\gamma,{\rm min}}^{\rm ph}/(1+z) > 50\;$GeV for the other time-bins.
Since the target photon field is highly anisotropic, the real break energies may be much larger than
those values (see references listed above Equation~\ref{eq:gammagamma}),
so that we have neglected the pair-creation breaks in Figure~\ref{fig:spec_080916C}.
We confirm that the second-order UP emission is negligible for each time-bin. For the second to
fifth time-bin spectra, we have $\eta \gamma_c m_e c^2 \ll \gamma_c^2 \varepsilon_{{\rm up},c}$,
so that the KN effect suppresses the second-order UP emission. For the first time-bin, the 
second-order UP emission flux is just much smaller than the other emission components.
The parameters satisfy our assumptions of $r_i \gg r_f$ for the first time-bin (Eq.~\ref{eq:cond_low}) and
$r_i \gg r_{\rm ph}$ for the other time-bins (Eq.~\ref{eq:cond_high}).

We have shown that the parameter regime should shift from $\eta > \eta_*$ into $\eta < \eta_*$
to reproduce the observed first and second time-bin spectra in our model. This shift is related to
the large decrease of $r_a/\Gamma_a$ from $\sim 2 \times 10^8\;$cm in the first time-bin to 
$\sim 1 \times 10^7\;$cm in the second 
time-bin. At this transition, the luminosity ratio of the photospheric and UP components 
$x = L_{\rm up}/L_{\rm ph}$ (Equations~\ref{eq:x_low} and \ref{eq:x_high}) increases as 
$r_a/\Gamma_a$ decreases (and then $\eta_*$ increases) and possibly $\epsilon_d \epsilon_e$ increases
because of the increasing differences of the final Lorentz factors of the shells. 
As we discuss in Section~\ref{sec:summary}, the simulations of the jet dynamics in the progenitor star
\citep{morsony07} suggest that $r_a/\Gamma_a$ may undergo a sudden decrease as time progresses.
Thus this transition may be interpreted as the reason for the observed delay of the LAT emission onset.
After the transition, the parameter sets constrained for the allowed smallest luminosities indicate that
$L$, $\eta$, and $t_v$ evolve monotonically, while 
$r_a/\Gamma_a$, $\mathcal{R}$, $\beta_{\rm ph}$, and $p$ are stable (note that the values of
$\epsilon_d \epsilon_e$ and $a_L$ are given to determine the other parameters).

The variability timescale is as small as $t_v \sim 10^{-3}\;$s, which is larger
than the light crossing timescale across a Schwarzschild black hole of
mass $\sim 10\;M_{\odot}$, $\sim 10^{-4}\;$s.
Such a timescale ($t_v(1+z) \sim 5\;$ms in the observer-frame) may not be resolved by the 
$\gamma$-ray detectors that observed this burst as far as we know.
The most sensitive detector among them for the $100\;{\rm keV} - 30\;$MeV range is INTEGRAL,
which has a resolution down to $50\;$ms and found a variability timescale $\sim 100\;$ms
\citep{greiner09}. An unsteady mass accretion from a convectively unstable torus onto the 
central compact object might produce such a large timescale of the flux change \citep{sekiguchi10}, 
or the interaction of the jet with the stellar envelope could induce a variability with light crossing 
timescale from the side to the
axis of the jet $\sim R_* \theta_j (1+z)/c \sim 100\;(R_*/5\times10^{10}\;{\rm cm}) (\theta_j/0.01)\;$ms,
where $R_*$ is the stellar radius and $\theta_j$ is the opening angle of the jet \citep{morsony10}.

As seen in Figure~\ref{fig:spec_080916C}, the low-energy portions of the
spectra of the photospheric emission are much harder than the observed spectral portions below 
the peak energies.  This is currently a generic problem for all photospheric mission models, 
which will be discussed in Section~\ref{sec:summary}. 

Interestingly, the observed peak energies and peak $\varepsilon F_{\varepsilon}$ fluxes in the 
second to fifth time-bins are clearly consistent with the famous peak energy-luminosity relation 
\citep[so-called Yonetoku relation;][]{yonetoku04,ghirlanda05,ghirlanda10}, which implies 
$\varepsilon_{\rm ph} \propto L_{\rm ph}^{1/2}$ in our model. This translates into a relation
$\eta \propto L^{7/16} (r_a/\Gamma_a)^{1/8} \mathcal{R}^{1/4}$ by Equations (\ref{eq:ph_high_baryon}).
Considering that $r_a/\Gamma_a$ and $\mathcal{R}$ are not much changed in our results and the 
dependences in the relation are weak, the dominant relation should be $\eta \propto L^{7/16}$ 
(or roughly $\eta \propto L^{1/2}$). A degree of fine tuning appears needed to satisfy this relation;
this is similar to other GRB emission models, which also require fine tunings of the parameters
of the jet \citep[e.g.,][]{ghirlanda10,rees05,thompson07,ioka10}.

Furthermore, we should clarify how tightly the parameter values related to the UP component
are constrained for reproducing the smooth Band-like spectrum by a superposition of the two components 
for each time-bin. 
Taking the second time-bin as an example, we make different model spectra by fixing $L$, $r_a/\Gamma_a$, 
$(\eta/\eta_*)^{8/3}$, and $\beta_{\rm ph}$ to have the same photospheric component, while changing the 
other parameters to see whether we obtain a smooth Band-like spectrum. The left model in 
Figure~\ref{fig:spec_dp} is the result for a value of $t_v$ that is $2$ times larger than the value in 
Table~\ref{tab:parameters}, which is marginally consistent with the data at $1\sigma$ level. 
Even larger $t_v$, which leads to larger $\gamma_c$ (and then larger $\varepsilon_{{\rm up},c}$ and 
smaller $h$), makes the UP component dimmer and then the overall spectrum significantly deviates from 
the data. We also examine models for smaller $\mathcal{R}$, which leads to smaller $\gamma_c$ (with
larger $\gamma_m$). This case keeps $h \sim 1$ (and then $L_{\rm up} = L \epsilon_d \epsilon_e h \sim 
{\rm const.}$), making the overall model spectrum somewhat bumpy. For a value of $\mathcal{R}$ that is 
$2$ times smaller than the value in Table~\ref{tab:parameters}, the bumpy model spectrum is marginally 
consistent with the data at $1\sigma$ level at $\varepsilon/(1+z) < 1\;{\rm GeV}$ but does not reproduce the 
flux at $\varepsilon/(1+z) > 1\;$GeV because of smaller $\varepsilon_{{\rm up},c}$. Even smaller $\mathcal{R}$ 
leads to the fast-cooling case $\gamma_c < \gamma_m$. The spectrum below the UP peak energy is 
harder than the slow-cooling case with $p \simeq 2.8$, so that the model spectrum is more bumpy,
although a higher $\varepsilon_{{\rm up},m}$ can explain the emission at the high-energy range.
The right model in Figure~\ref{fig:spec_dp} is the result for a value of $\mathcal{R}$ that is 
$5$ times smaller than the value in Table~\ref{tab:parameters}, which is still marginally consistent
with the data at $1\sigma$ level. For a smaller $\mathcal{R}$, however, the overall model spectrum
significantly deviates from the data in the middle range $\sim 10\;{\rm MeV} - 1\;{\rm GeV}$.
This parametric study indicates that the parameter spaces of $t_v$ and $\mathcal{R}$ are limited within 
a factor of $\sim$ a few for obtaining a Band-like function by the superposition of the two components. 

\begin{figure*}
\begin{center}
\includegraphics{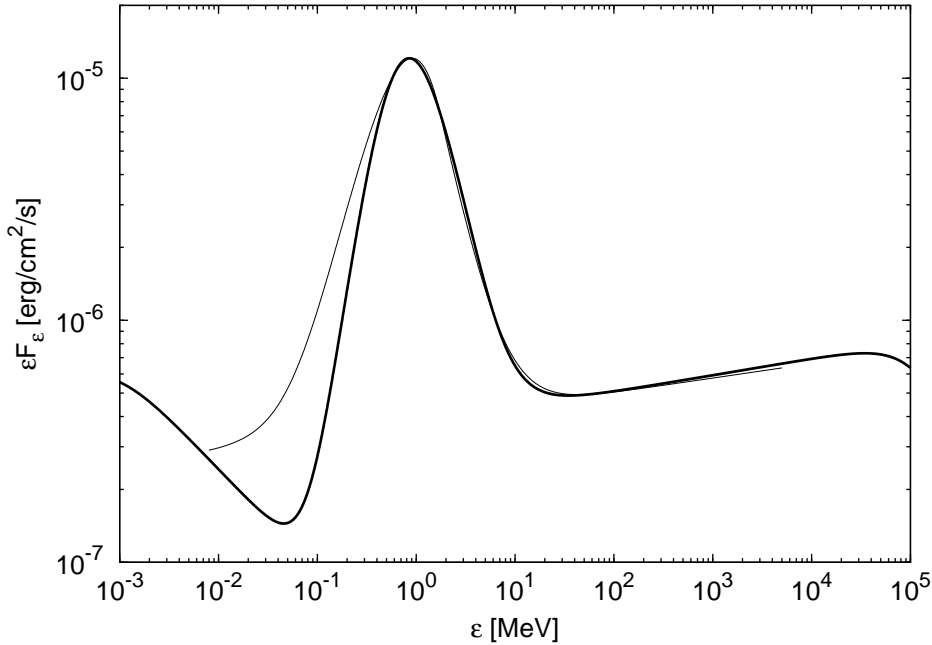}
\end{center}
\caption{
Model fit of the spectrum of GRB 090902B in the $4.6-9.6\;$s time bin. The thin line is the observed data,
and the thick line is the model fit. Adopted parameters are listed in Table~\ref{tab:parameters}.
The photospheric and UP emission components are 
consistent with the observed spectra of the Band component at and above the peak energy
and the distinct high-energy power-law component, respectively. 
The observed low-energy spectrum of the Band component is softer than the model. This issue is
discussed in Section~\ref{sec:summary}.
A deviation from the Band component at $\la 40\;$keV could be explained by a contribution from 
the synchrotron component.}      
\label{fig:spec_090902B}
\end{figure*} 

This means that in order to obtain the Band-like spectra through the second to fifth time-bins, we need 
significant fine tuning of the model parameters $t_v$ and $\mathcal{R}$ (as well as $\beta_{\rm ph}$ 
and $p$). However, we note that \citet{abdo080916C} conclude ``Compared to the null hypothesis that 
the data originated from a simple Band GRB function, adding the additional power-law component resulted
in a probability of 1\% that there was no additional spectral component for this (fourth) time bin.''
Taken at face value, this would imply that the fourth time-bin data may have 
an additional high-energy spectral component besides the Band function although the significance level
is low ($\sim 2\sigma$ level). Furthermore, the light curve at $< 10^3\;$s in \citet{abdo080916C} shows
the abrupt steepening break of the GBM light curve at the fifth time-bin while the LAT flux decays stably, 
which implies a two component origin at this time-bin. \citet{binbin10} claim that the background 
uncertainty significantly affect the results of the spectral analysis due to the low count rate at 
such late times. These implications increase the plausibility of our two component model.

\subsection{GRB 090902B}

This burst is a long GRB that occurred at a redshift $z \simeq 1.82$ (which corresponds to 
$d_L \simeq 4.3 \times 10^{28}\;$cm). The results of the time-resolved spectral analysis 
by \citet{abdo090902B} show that all the spectra can be fitted by a Band
function plus a distinct power-law function. They did not list the flux normalizations of the 
two components and only show an overall spectrum for the second time-bin $4.6-9.6\;$s, which
we show by the thin line in Figure~\ref{fig:spec_090902B}. We apply our model only
for this time-bin spectrum.
The LAT emission starts $\sim 3\;$s after the GBM trigger within the first time-bin $0.0-4.6\;$s.
The delay time in the cosmological rest frame is $\sim 3/(1+z)\;{\rm s} \sim 1\;$s, which is similar to
that of GRB 080916C.

\begin{figure*}
\begin{center}
\includegraphics{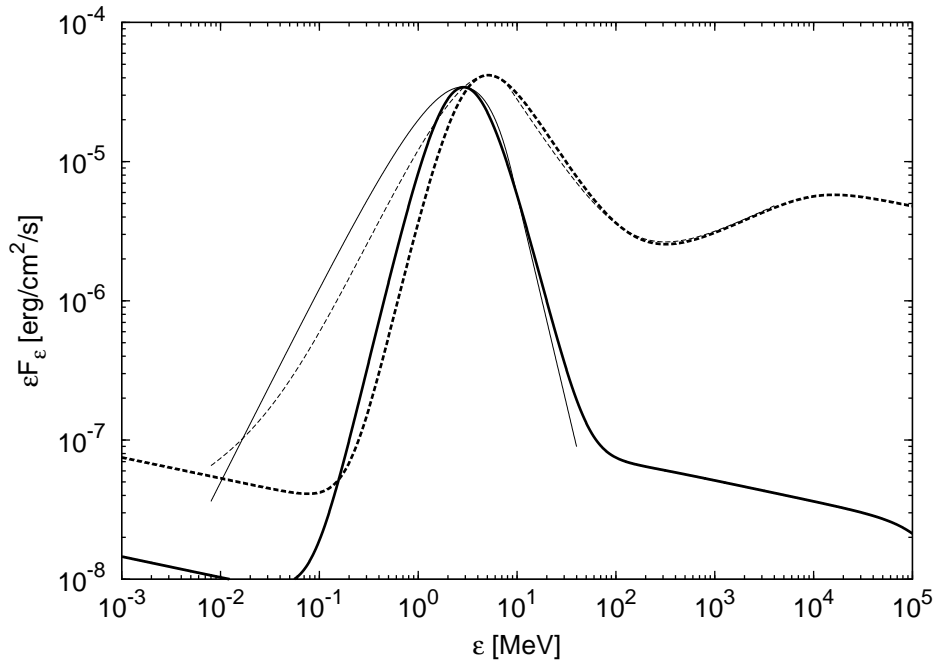}
\end{center}
\caption{
Model fits of the time-resolved spectra of GRB 090510. The thin lines are the observed data,
and the thick lines are the model fits (the solid lines are for $0.5-0.6\;$s, 
and the dashed lines are for $0.6-0.8\;$s). 
Model parameter values are listed in Table~\ref{tab:parameters}. 
The observed spectral parts below the peak energies are significantly softer than the model.
This issue is discussed in Section~\ref{sec:summary}.}
\label{fig:spec_090510}
\end{figure*}

The Band and distinct high-energy power-law components of the second time-bin spectrum 
can be straightforwardly modeled by the photospheric and UP emission, respectively, as shown
by the thick line in Figure~\ref{fig:spec_090902B}. 
The observed data of {\it Fermi}/GBM show a clear deviation from the Band
function at $\la 40\;$keV, and the thin line is obtained by assuming that the low-energy excess 
and the high-energy emission at $\ga 50\;$MeV are the same power-law emission component \citep{abdo090902B}.
In our model, however, the low-energy excess could be a contribution from the synchrotron component, which 
is separate from the high-energy UP component. If the superposition of the multiple photospheric emission 
from different shells can reproduce the observed low-energy power-law portion of the Band component (see 
Section~\ref{sec:summary} for more discussion) and it extends even below $\sim 40\;$keV, the contribution 
from the synchrotron component could explain the observed low-energy excess.
The model fit is obtained by setting the conditions
$L_{\rm ph} \sim 2.8 \times 10^{53}\;{\rm erg}\;{\rm s}^{-1}$, $\varepsilon_{\rm ph} \sim 2.4\;$MeV,
$\beta_{\rm ph} \sim -4.0$, $L_{\rm up} \ga 1.3 \times 10^{52}\;{\rm erg}\;{\rm s}^{-1}$,
$\varepsilon_{{\rm up},c} \ga 40\;$GeV, $\varepsilon_{{\rm up},m} \la 16\;$MeV, and $p \sim 2.9$
(which correspond to $\gamma_c \ga 130$ and $\gamma_m \la 2.6$).
For these values we have $L_{\rm up}/L_{\rm ph} = x \ga 0.06$, 
and $h = (\gamma_c/\gamma_m)^{2-p}/(3-p) \la 0.2$. 
These lead to $\eta < \eta_*$ for a reasonable assumption of $\epsilon_d \epsilon_e \la 0.1$, and thus 
this spectrum corresponds to case 1, similar to the second time-bin of GRB 080916C.
From these conditions and given values $\epsilon_d \epsilon_e \sim 0.1$ and $a_L \sim 5$, 
we can constrain the model parameter values, in a similar way to that for GRB 080916C.
A larger $h$ (for a smaller $\gamma_c$ and a larger $\gamma_m$) leads to the 
smaller luminosity $L$, while in order to explain the low-energy excess as the synchrotron emission at
$\varepsilon > \varepsilon_{{\rm syn},c}$ with a reasonable range $\epsilon_d \epsilon_B \la 0.1$, 
we require $\gamma_c \ga 350$ (for which $L_{\rm up} \ga 1.6 \times 10^{52}\;{\rm erg}\;{\rm s}^{-1}$ 
is needed to explain the high-energy emission).
In Table~\ref{tab:parameters}, we show the parameter sets constrained for the case of $\gamma_c \sim 350$,
$\gamma_m \sim 2.6$, and $L_{\rm up} \sim 1.6 \times 10^{52}\;{\rm erg}\;{\rm s}^{-1}$.
They satisfy our assumption of $r_i \gg r_{\rm ph}$ (Eq. \ref{eq:cond_high}). 
The internal shock region is found to be optically thin for $e^{\pm}$ pair creation.
The second-order UP emission is confirmed to be negligible.

It has been reported that the Band component of this burst can be fitted by a multi-temperature 
blackbody spectrum \citep{ryde10}. The observed low-energy spectral slope of the Band component 
$\alpha \simeq 0.07$ is relatively hard and near (but still softer than) the single-temperature
thermal model $\alpha_{\rm ph} = 1$, and the high-energy spectral slope $\beta \simeq -3.9$ is 
very steep. These strongly favor a photospheric origin of the Band component of this burst. 

The first time-bin of GRB 090902B includes a very small number of the high-energy photons,
which implies larger $\eta/\eta_*$. Thus the LAT onset delay in this burst may be consistent with
the interpretation that the parameter regime shifts from $\eta > \eta_*$ in the first time-bin
into $\eta < \eta_*$ in the second time-bin, similar to the case of GRB 080916C.

\subsection{GRB 090510}

\begin{figure*}
\begin{center}
\includegraphics{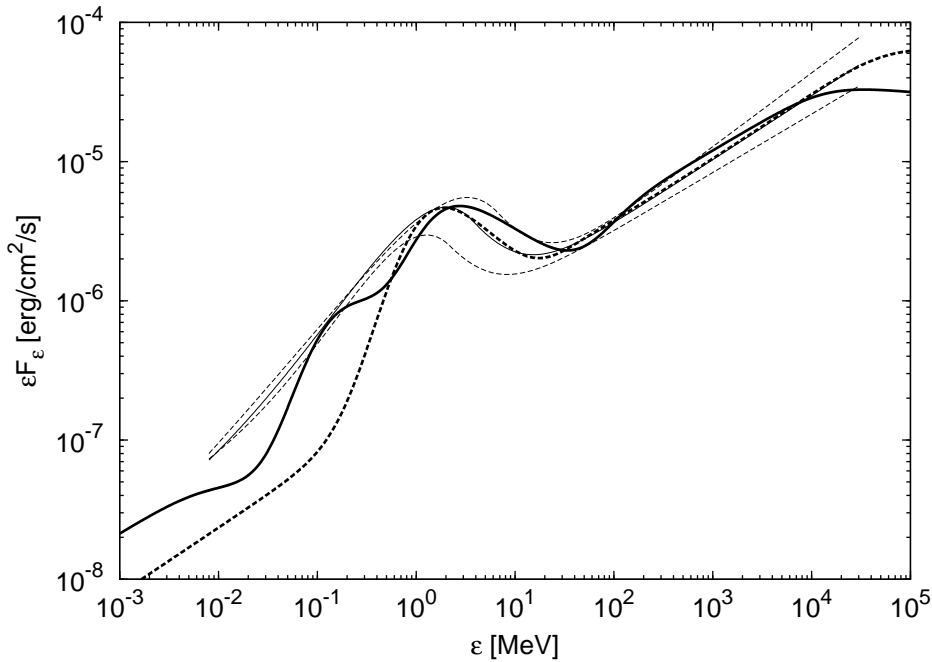}
\end{center}
\caption{
Model fits of the third time-bin ($0.8-0.9\;$s) spectrum of GRB 090510. The thin line is the Band plus
power-law function fitting to the observed data and the thin dashed lines represent the $1\sigma$ errors.
The thick solid and dashed lines are our model fits. Model parameter values are listed in 
Table~\ref{tab:parameters}. See texts for details.}      
\label{fig:spec_090510_3}
\end{figure*}

This burst is a short GRB that occurred at a redshift $z \simeq 0.90$ (corresponding to 
$d_L \simeq 1.8 \times 10^{28}\;$cm). This has a precursor $\sim 0.5\;$s earlier than the main burst. 
The analysis results of four time-binned spectra of the main burst are given in
\citet{ackermann090510}.
The first and second time-bin spectra are shown by the thin lines in Figure~\ref{fig:spec_090510}, 
and the third time-bin spectrum in Figure~\ref{fig:spec_090510_3}.
The first time-bin spectrum can be fitted by a Band function only,
while the second and third time-bin spectra can be fitted by Band plus distinct power-law
functions. For the last time-bin $0.9-1.0\;$s, the emission was detected
only in the LAT energy range ($100\;{\rm MeV}-2\;{\rm GeV}$), whose spectrum is fitted by 
a power-law function. We will not fit this last time-bin spectrum with our model but we 
discuss a possible explanation for it in the context of our model.
The onset delay of the distinct high-energy component with respect
to the Band component in the cosmological rest frame is estimated to be $\sim 0.1/(1+z)\;{\rm s} \sim 0.05\;$s.
It is remarkable that the distinct high-energy component is brighter than the Band component 
in the third time-bin, which is unique among all the observed LAT GRBs.

We first consider the {\it second} time-bin spectrum. This may be explained by a combination of 
the photospheric and UP emission components. The model parameters for this spectrum can be found in
a similar way to the case of GRB 090902B. The luminosity ratio is required to be 
$L_{\rm up}/L_{\rm ph} = x \ga 0.1$,
and we estimate that $h \la 0.7$. These lead to $\eta \la \eta_*$ for a reasonable 
assumption $\epsilon_d \epsilon_e \la 0.1$, and thus this spectrum corresponds to case 1, 
similar to the second time-bins of GRB 080916C and GRB 090902B. 
The best fit model for this spectrum is shown by the thick dashed line in 
Figure~\ref{fig:spec_090510}, for which
$L_{\rm ph} \sim 1.7 \times 10^{53}\;{\rm erg}\;{\rm s}^{-1}$, $\varepsilon_{\rm ph} \sim 9.7\;$MeV,
$\beta_{\rm ph} \simeq -3.0$, $L_{\rm up} \ga 2.3 \times 10^{52}\;{\rm erg}\;{\rm s}^{-1}$, 
$\varepsilon_{{\rm up},c} \ga 20\;$GeV, $\varepsilon_{{\rm up},m} \la 44\;$MeV,
and $p \simeq 2.3$ are required. 
From these conditions and given values $\epsilon_d \epsilon_e \sim 0.1$ and $a_L \sim 5$, 
we can constrain the model parameter values in a similar way to the cases for GRB 080916C and GRB 090902B.
The parameter sets constrained for the allowed smallest luminosity 
(i.e., for $\varepsilon_{{\rm up},c} \sim 20\;$GeV, $\varepsilon_{{\rm up},m} \sim 44\;$MeV, which lead
to the largest $h$, and $L_{\rm up} \sim 2.3 \times 10^{52}\;{\rm erg}\;{\rm s}^{-1}$) are shown 
in Table~\ref{tab:parameters}.

Next we consider the {\it first} time-bin spectrum. This should be produced by the photospheric
emission only and the luminosity ratio is required to be $L_{\rm up}/L_{\rm ph} = x \la 3\times10^{-3}$. 
For a reasonable value $\epsilon_d \epsilon_e \sim 0.1$, we have $\eta > \eta_*$. 
This spectrum corresponds to case 9 (see Table~\ref{tab:cases}).
The model fit shown by the thick solid line in Figure~\ref{fig:spec_090510} requires that the photospheric
emission has $L_{\rm ph} \sim 1.4 \times 10^{53}\;{\rm erg}\;{\rm s}^{-1}$, $\varepsilon_{\rm ph} 
\sim 5.5\;$MeV, and $\beta_{\rm ph} \sim -4.8$. Equations (\ref{eq:ph_low_baryon}) determine 
$L_{53} \simeq L_{{\rm ph},53} \sim 1$ and $r_{a,7}/\Gamma_a \sim 3$. These lead to a very large
$\eta/\eta_* \simeq L \epsilon_d \epsilon_e h/L_{\rm up} \sim 30 (\epsilon_d \epsilon_e/0.1)$, indicating
a very large $\eta \sim 10^5$. The parameters for the model fit shown in Figure~\ref{fig:spec_090510}
are shown in Table~\ref{tab:parameters}.

We can also consider a case of $\eta > \eta_*$ and $\epsilon_d \epsilon_e \ll 0.1$ for the first 
time-bin, similar to the first time-bin of GRB 080916C. We show a result for a case of 
$\epsilon_d \epsilon_e \sim 5 \times 10^{-3}$ in Table~\ref{tab:parameters}, in which 
$\eta/\eta_* \sim 2$ and $\eta \simeq 5.4 \times 10^3$. 
In this case Equation~(\ref{eq:gamma_m}) provides $\gamma_m \sim 0.1 < 1$, which means in reality 
that most of the leptons have $\gamma \sim 1$ and only a small fraction $\sim 0.1$ of them participate
in the non-thermal power-law acceleration. 
The UP component spectrum has a bump around $\varepsilon_{\rm ph}$, which makes a small contribution
to the overall spectrum.

Next we find the parameter values appropriate for the third time-bin spectrum. In this time-bin
the UP luminosity has to be higher than the photospheric luminosity. This may correspond to case 2
(see Table~\ref{tab:cases}), unlike the cases considered above. 
The spectrum can be fitted as shown by the thick dashed line in Figure~\ref{fig:spec_090510_3},
where the emission component at $\varepsilon/(1+z) < 0.1\;$MeV is the SSC emission (the second-order SSC 
emission is hidden by the UP emission).
The characteristic quantities required for this model fit are $L_{\rm ph} \sim 1.8 \times
10^{52}\;{\rm erg}\;{\rm s}^{-1}$, $\varepsilon_{\rm ph} \sim 3.6\;{\rm MeV}$, $\beta_{\rm ph} \simeq -2.8$,
$L_{\rm up} \ga 2.3 \times 10^{53}\;{\rm erg}\;{\rm s}^{-1}$, $\varepsilon_{{\rm up},c} \ga 100\;$GeV, 
$\varepsilon_{{\rm up},m} \la 50\;$MeV, and $p \simeq 2.05$. 
For a somewhat larger $\epsilon_d \epsilon_e \sim 0.15$, we can constrain the model parameter values
in a similar way to case 1, and the results are shown in Table~\ref{tab:parameters}.
The value $r_{a,7}/\Gamma_a \sim 2 \times 10^{-3}$ is much smaller than that in the second time-bin.
This leads to $r_a \ga 2 \times 10^4\;$cm, which is too small, compared to a Schwarzschild radius 
$\sim 3 \times 10^5\;$cm of the central compact object of mass $\sim M_\odot$.
If we take a smaller $L_{\rm up}$ and $\varepsilon_{{\rm up},c}$ which are consistent with the data
at $1\sigma$ level, i.e., $L_{\rm up} \sim 1.2 \times 10^{53}\;{\rm erg}\;{\rm s}^{-1}$ and 
$\varepsilon_{{\rm up},c} \sim 25\;$GeV, we have $r_a/\Gamma_a \sim 6 \times 10^4\;$cm (and
we have $L_{53} \sim 10$ smaller than the larger $L_{\rm up}$ model). 
At such a late phase of the prompt emission, the external shock emission might contribute to the 
high-energy emission \citep{depasquale10}, but this possibility is suggested to be unlikely
\citep{he10} \citep[see also][]{liu10}.
Below we show another sets of parameter values which could explain the third time-bin emission
at $1\sigma$ level together with the fourth time-bin emission, and have an even larger $r_a/\Gamma_a$.

Finally we discuss the fourth time-bin spectrum. In this time-bin, only the high-energy photons
are detected, which requires much larger $\epsilon_d \epsilon_e$ and/or much smaller $r_a$.
These do not seem realistic. Here we show an example of another parameter set for the {\it third}
time-bin spectrum, for which the duration of the UP emission can be larger than the pulse width
of the MeV photospheric emission and stays bright even in the fourth time-bin without the MeV emission.
Here we consider the inefficient scattering regime for the third time-bin (see Section~\ref{sec:temporal};
we have assumed the efficient scattering regime for the first and second time-bins, say a case of
$\tilde{W} \sim c \tilde{t}_v$), in which the widths of the two shells $W$ are not much different from 
those for the earlier time-bins $\tilde{W}$, but the separation of the two shells $c t_v$ is much larger 
than those for the earlier time-bins $c \tilde{t}_v$. In this regime we consider a case of 
$a_L^{-2} c t_v \sim W + c \tilde{t}_v \sim 2 c \tilde{t}_v$, for which the leptons in the 
internal shock of the given two shells with $W$ and $t_v$ up-scatter the photospheric emission from 
a third shell at a distance $c \tilde{t}_v$ behind these two shells.
In this case the angular spreading timescale of the UP emission is estimated to be
$\sim 3 a_L^{-1} t_v$, which is much larger than $W/c$ and $\tilde{t}_v$, so that we have 
$\delta t_{\rm up} \sim 3 a_L^{-1} t_v$ (see Eq.~\ref{eq:up_dur}).
Since $\delta t_{\rm ph} \sim W/c$, which means that the UP emission
can last much longer than the MeV emission. If $\delta t_{\rm up} (1+z)$ is comparable
or larger than the duration of the time-bin, $0.1\;$s, this scenario could explain the fourth time-bin
spectrum. Below we show that the model fit of the third time-bin
spectrum in this scenario provides a $t_v$ consistent with the above temporal conditions.

We interpret the observed Band and distinct high-energy components in the third time-bin as the 
photospheric emission of the third shell and the UP emission produced by up-scattering of the 
photospheric emission of the third shell by the internal shock of the first and second shells, respectively. 
The model fit is shown by the thick solid line in Figure~\ref{fig:spec_090510_3}, which is marginally
consistent with the data at $1\sigma$ level.
The photospheric emission of the third shell has $\tilde{L}_{\rm ph}
\sim 1.8 \times 10^{52}\;{\rm erg}\;{\rm s}^{-1}$, $\tilde{\varepsilon}_{\rm ph} \sim 5.5\;$MeV, and
$\tilde{\beta}_{\rm ph} \simeq -2.5$. (Hereafter a tilde denotes a quantity of the third shell.) 
The UP emission is required to have $L_{\rm up} \simeq L \epsilon_d \epsilon_e h \sim 1.2 \times 10^{53}\;
{\rm erg}\;{\rm s}^{-1}$, $\varepsilon_{{\rm up},c} \simeq \tilde{\varepsilon}_{\rm ph} \gamma_c^2
\sim 25\;$GeV, $\varepsilon_{{\rm up},m} \simeq \tilde{\varepsilon}_{\rm ph} \gamma_m^2
\sim 180\;$MeV, and $p \simeq 2.15$. The condition for $L_{\rm up}$ gives us $L_{53} \sim 10$ 
for a given $\epsilon_d \epsilon_e \sim 0.15$. 
The photospheric flux of the second shell should be lower than the observed flux level. 
We take $L_{\rm ph} \sim 3.5 \times 10^{51}\;{\rm erg}\;{\rm s}^{-1}$
and $\varepsilon_{\rm ph} \sim 0.5\;$MeV which produces the spectral bump below the peak energy
of the main photospheric emission in Figure~\ref{fig:spec_090510_3} and requires
$r_{a,7}/\Gamma_a \sim 0.01$ and $(\eta/\eta_*)^{8/3} \sim 4 \times 10^{-3}$ through 
Equation~(\ref{eq:ph_high_baryon}). Equation~(\ref{eq:gamma_m}) leads to $\mathcal{R}_1 \sim 0.6$.
Then we have $\eta_{*,3} \sim 13$ and $\eta_3 \sim 1.6$.
Finally the remaining parameter $t_v$ is determined by an equation for $\gamma_c$, which is given by
Equation~(\ref{eq:gamma_c_high}) multiplied by a factor $L_{\rm ph}/\tilde{L}_{\rm ph} \sim 0.2$. 
This factor leads to $t_{v,-2} \sim 7$, much larger than in the above simple two-shell scenario. 
The observed duration of the 
UP emission pulse is then estimated to be $\delta t_{\rm up} (1+z) \sim 3 a_L^{-1} t_v (1+z) \sim 0.08\;$s. 
This is comparable to the durations of the third and fourth time-bins and roughly consistent with the condition 
$a_L^{-2} c t_v \sim 2 c \tilde{t}_v $, so that the high-energy emission without a corresponding MeV emission
in the fourth time-bin could be explained by the UP emission with large angular spreading time.

The value $r_a/\Gamma_a \sim 10^5\;$cm is comparable but still somewhat smaller than a Schwartzschild 
radius of mass $\sim M_\odot$. A larger $r_a/\Gamma_a$ would lead to a brighter spectral bump of the 
photospheric emission from the second shell at $\sim 0.1\;$MeV in Figure~\ref{fig:spec_090510_3}, as 
$L_{\rm ph} \propto (r_a/\Gamma_a)^{2/3}$ and 
$\varepsilon_{\rm ph} \propto (r_a/\Gamma_a)^{1/6}$ in Eq.~(\ref{eq:ph_high_baryon}),
violating the observed data. This might suggest that a large fraction 
of the jet energy at $r_a$ is not thermal but Poynting flux for this burst \citep[cf.][]{zhang09}.

To summarize this scenario, the pulse widths of the UP component 
$\delta t_{\rm up} \sim W/c + 3 a_L^{-1} \tilde{t}_v$ are similar to those of the photospheric component 
$\delta t_{\rm ph} \sim W/c$ (as $\sim 1\;$ms) at the first and second time-bins,
while at the third time-bin the UP pulse width only increases significantly to 
$\delta t_{\rm up} \sim 3 a_L^{-1} t_v \sim 40\;$ms, which may stay bright
even at the fourth time-bin. This may be consistent with the results of the cross-correlation function 
analysis by \citet{ackermann090510} that there is no correlated variability between the GBM and LAT 
emission.\footnote{Nevertheless, we can see some narrow spikes of the LAT emission correlated 
with GBM pulses throughout the whole prompt phase.
This may not be inconsistent with our second (large $\delta t_{\rm up}$) scenario for the third time-bin,
since the third time-bin model spectrum indicates that the high-energy tail part of the photospheric 
emission which has sharp variability contributes to $\sim 50\%$ of ($\sim 10\%$ of) the observed 
photon numbers at the entire LAT range $\ga 20\;$MeV (at $>100\;$MeV). Also, the UP emission
comes from a large area of the shell with solid angle of $\la 3/\Gamma$ (see Section~\ref{sec:temporal}), 
which could have angular inhomogeneity leading to the sharp variability at the both third and fourth 
time-bins.} 
In this scenario there should be a large interval $t_v (1+z) \sim 0.1\;$s of the photospheric emission
before the third time-bin, which seems consistent with the observed quiescent time in the 
$260\;{\rm keV} - 5\;$MeV light curve.
Such a variability timescale $\sim 0.1\;$s might arise from the mass accretion on the central
compact object from an inhomogeneous torus \citep{rosswog07} or from the interaction of the jet
with the dense environment \citep{morsony10}.

The parameter sets we found for the three time-bin spectra are summarized in Table~\ref{tab:parameters}.
We have shown that the parameter regime should shift from $\eta > \eta_*$ in the first time-bin into 
$\eta < \eta_*$ in the second time-bin, which is related to the large decrease of $r_a/\Gamma_a$ 
and corresponds to the increase of the luminosity ratio $x = L_{\rm up}/L_{\rm ph}$ and the delayed
onset of the distinct high-energy component, similar to the cases of GRB 080916C and GRB 090902B.
The second and third time-bin spectra satisfy the relation $\varepsilon_{\rm ph} \propto L_{\rm ph}^{1/2}$,
similar to the case of GRB 080916C, which requires a fine tuning of the parameters $L$, $r_a/\Gamma_a$,
$\eta$, and $\mathcal{R}$.

The synchrotron and SSC fluxes should be lower than the observed flux levels, which put constraints on
$\epsilon_d \epsilon_B$, which is much tighter than the cases of GRB 080916C and GRB 090902B. 
In particular for the third time-bin, we need $\epsilon_d \epsilon_B \la 3 \times 10^{-6}$.
A typical value $\epsilon_d \sim 0.25$ leads to a constraint $\epsilon_B \la 10^{-5}$. This is 
not implausible since some external shocks driven by GRBs have been suggested to have such 
small values of $\epsilon_B$ \citep{panaitescu02,granot03,peer04}.
The internal shock region is estimated to be optically thin for the $e^{\pm}$ pair creation for the 
second and third time-bins, and we find $\varepsilon^{\rm min}_{\gamma\gamma,{\rm ph}}/(1+z) \sim 100\;$GeV
for the first time-bin.
The parameter sets for all the time-bins satisfy our assumptions of Equations (\ref{eq:cond_high})
and (\ref{eq:cond_low}).
The second-order UP emission is negligible for all the time-bins.

\section{Summary and Discussion}
\label{sec:summary}

GRB jets are thought to consist of many successive shells moving at relativistic speeds. These 
naturally lead to variable photospheric emission around the MeV energy range, and also to 
internal shocks above the photosphere. 
Energy dissipation near the photosphere caused by e.g., internal shocks, excited plasma waves, 
and/or interaction of the jet with the dense environment, or nuclear collisions between protons
and neutrons is expected to make the photospheric emission having the power-law tail spectrum
above the peak energy $\varepsilon_{\rm ph}$, with the photon index $\beta_{\rm ph} \la -2.5$ 
under the assumption of the roughly adiabatic evolution of the jet (see Section~\ref{sec:ph}).
We have generically studied the temporal and spectral properties of the radiation from the photosphere
and the internal shock in the jet which is accelerated by the thermal pressure.
We have not considered acceleration by processes related to magnetic fields.
We have shown that the photospheric emission is efficiently up-scattered 
to the high-energy range by the electrons (and positrons) accelerated in the internal shocks, for the
efficient scattering regime $(\Gamma_s^2/\Gamma_r^2)t_v < W/2$ which includes the 
typical case $W \sim c t_v$ (see Section \ref{sec:temporal}), when the radiation from the internal 
shocks consist of the UP, synchrotron, and SSC emission components. 

Our generic arguments show that the ordering of the luminosities of the emission components
depends on the values of $(\eta/\eta_*)^{8/3}$, $\epsilon_d \epsilon_e h$, and $\epsilon_d \epsilon_B$ 
(see Section~\ref{sec:spectral} and Table~\ref{tab:cases}). 
A condition $(\eta/\eta_*)^{8/3} \gg {\rm max}(\epsilon_d \epsilon_e h, \epsilon_d \epsilon_B)$ is
required in order to have the photospheric and UP components dominant in the MeV energy range and 
in the high-energy range, respectively, rather than the synchrotron and SSC components, and 
$L_{\rm ph} \gg L_{\rm up}$ as typically observed in LAT GRBs. (We can also have a case of
$L_{\rm up} \gg L_{\rm ph}$ for a condition $\epsilon_d \epsilon_d h \gg (\eta/\eta_*)^{8/3} \gg
[(\epsilon_d \epsilon_e)^2 \epsilon_d \epsilon_B]^{1/3}$, which has been applied to the third time-bin
spectrum of GRB 090510.) For reasonable values of 
$\epsilon_d \epsilon_e$ and $\epsilon_d \epsilon_B$ both $\la 0.1$ for internal shocks, that 
condition reduces to $\eta \ga \eta_* \simeq 2.8 \times 10^3 L_{53}^{1/4} 
(r_{a,7}/\Gamma_a)^{-1/4} \mathcal{R}_1^{1/4}$ (Equation~\ref{eq:eta_*}).
In this case the photospheric emission has luminosities $L_{\rm ph} \sim L$, and its peak energy is 
given by $\varepsilon_{\rm ph} \sim 8\; L_{53}^{1/4} (r_{a,7}/\Gamma_a)^{-1/2}\;$MeV 
(Equations~\ref{eq:T_a}, \ref{eq:ph_low_baryon}, and \ref{eq:ph_high_baryon}), which can be 
consistent with the Band components of the observed LAT GRBs as well as other GRBs
since we can have $10^{-1} \la r_{a,7}/\Gamma_a \la 10^2$ in principle
(but there are several issues to be addressed on the spectral shapes and their temporal evolution; see below).
This implies that the GRB jets should typically have a relatively large bulk Lorentz factor at the 
prompt emission site, $\sim \eta_*$, in our model.

The UP emission component is a good candidate for the observed high-energy emission of LAT GRBs in the 
prompt phase. The long-lived high-energy emission after the prompt phase may be related 
to the external shock \citep[e.g.,][]{depasquale10,kumar10,ghisellini10,he10,liu10}.
We have performed case studies of the prompt emission of GRB 080916C, GRB 090902B,
and GRB 090510, and fitted their time-binned spectra by our model typically
with reasonable sets of parameter values (see Section~\ref{sec:case} and Table~\ref{tab:parameters}).
For GRB 090902B and GRB 090510, 
the UP emission corresponds to the observed distinct high-energy component, while for GRB 080916C, 
the UP emission is dominant in the high-energy range but the photospheric plus UP emission mimic the 
Band function between $\sim 10\;$keV and $\sim 10\;$GeV. 

We note that fine tuning of the parameter values appears to be required for some spectra in addition
to keeping $\eta \ga \eta_*$. 
In order for the combined photospheric and UP components to mimic a simple Band 
function fit of the spectra of GRB 080916C through the second to fifth time-bins, significant
fine tuning of several parameters is required.
Such a composite model is, however, compatible with the presence of an additional high-energy component, 
which is weakly suggested in the data of the fourth time-bin, and also the difference of the 
temporal behaviors of the GBM and LAT emission suggests a two component origin for the fifth time-bin.
Furthermore, the observed spectra of GRB 080916C and GRB 090510 show the temporal evolution roughly
obeying $\varepsilon_{\rm ph} \propto L_{\rm ph}^{1/2}$ from the second time-bins 
\citep[see also][]{ghirlanda10,yonetoku04}, which require a relation $\eta \propto L^{7/16}$.
For the third and fourth time-bins of GRB 090510, when the distinct
high-energy component is much brighter than the Band component, we require an additional
shell behind the two colliding shells and besides need $r_a/\Gamma_a \sim 10^5\;$cm close to or 
even smaller than a Schwartzschild radius of mass $\sim M_\odot$, $\sim 3 \times 10^5\;$cm, and very small 
$\epsilon_d \epsilon_B \la 3 \times 10^{-6}$. 

Other LAT GRBs, such as GRB 080825C \citep{abdo080825C} and GRB 081024B \citep{abdo081024B}, whose 
redshifts are not determined, display time-binned spectra fitted by Band functions with a temporal 
behavior of the high-energy spectral indices $\beta$ similar to those of GRB 080916C, so that they 
also may be explained by our model. In our model
the UP emission should exist in all GRBs if the jet is in the efficient scattering regime, but for
GRBs with small $L$ this may be below the detection threshold of LAT, since the UP luminosity is proportional 
to $L$. This is a general statement applicable to wide types of models in which $L$ determines the 
flux normalizations of both the Band and the high-energy emission components, and 
consistent with the fact that the high-energy emission is detected by LAT only in the 
brightest class of {\it Fermi} GRBs \citep{granot10}.

We have shown that if the GRB jet is in the efficient scattering regime right from the start,
the simple kinematic effect causes the onset of the first UP emission to be delayed with respect to that 
of the first photospheric emission by a timescale comparable to the pulse widths or pulse separations 
of the photospheric emission, which should be smaller than the variability timescale apparent in the 
MeV light curve (see Section~\ref{sec:temporal}). This may not explain the 
time delays much larger than the apparent variability timescale observed in most of the LAT GRBs.
In our model, such observed large time delays may instead be attributed to a temporal evolution of the 
jet parameters, $L$, $r_a/\Gamma_a$, $\eta$, $\mathcal{R}$, and $t_v$, a possible origin for which
is discussed in the next paragraph.
From the model fits of the time-binned spectra of the three LAT GRBs
(in Section~\ref{sec:case}), we found that the parameter regime should shift from $\eta > \eta_*$ into 
$\eta < \eta_*$ to reproduce the observed spectra (this shift leads to the increase of 
$L_{\rm up}/L_{\rm ph} = x$; Equation~\ref{eq:x_high} and \ref{eq:x_low}). 
In this case the UP emission component starts with a small delay with respect to the photospheric 
emission onset by the timescale comparable to the photospheric variability time,
being still dim compared with the photospheric emission while $\eta > \eta_*$, but starts to be
bright and detected by LAT when the parameters shift into the regime of $\eta < \eta_*$.

This parameter shift appears to be related to the large decrease of $r_a/\Gamma_a$, from $\sim 10^8\;$cm
to $\sim 10^7\;$cm, in the timescale of $t_{\rm delay} \sim 1\;$s in long GRBs GRB 080916C and 090902B 
(Table~\ref{tab:parameters}).
The progenitors of long GRBs are thought to be collapsing massive stars, and 
such a decrease of $r_a/\Gamma_a$ may be explained by the interaction of the jet with the stellar 
envelope before the jet breakout.
The numerical simulations by \citet{morsony07} \citep[see also][]{lazzati09} show that the jet
has three parts around the time when it breaks out the star; the jet head, the collimation-shocked part,
and the free expansion part. The jet head is defined as the very thin part at the front end of the jet 
between the forward and reverse shocks, which typically has a mildly relativistic speed and the material 
inside this part escapes sideways into the cocoon; a large fraction of the jet behind the jet
head, suffering weaker dissipation by collimation shocks, has a relativistic
speed; and behind the collimation-shocked part, the jet material suffers much weaker dissipation,
expanding adiabatically.\footnote{
For example, Figure~2 of \citet{lazzati09} shows that the collimation-shocked 
part has an averaged Lorentz factor of $\sim 10^2$ while the free 
expansion part which follows has a saturated Lorentz factor of $\eta \sim 400$.}
For the collimation-shocked part, $r_a$ may be the stellar radius
and $\Gamma_a$ is given by its Lorentz factor at the breakout time, while
for the free expansion part, $r_a$ may be the size of the central compact
object and $\Gamma_a \sim 1$. In such a jet structure, the emission properties may be changed at 
the transition between the collimation-shocked part and the free expansion part.
The observed time delay would then correspond to the length scale of the collimation-shocked portion 
of the jet at the breakout time, $l_d \sim c t_{\rm delay} \sim 3 \times 10^{10}\;$cm. The radius
of this part, which should be slightly larger than $l_d$, say $r_a \sim 5 \times 10^{10}\;$cm,
could be consistent with the size of the progenitor star and with our spectral modeling if 
$\Gamma_a \sim 500$, which is relativistic but smaller than $\eta \sim 3 \times 10^3$.
The value $r_a/\Gamma_a \sim 10^7\;$cm for the free expansion part is what might be expected from a 
compact object size $r_a \sim 10^7\;$cm and $\Gamma_a \sim 1$.
This $r_a$ is a few Schwarzschild radii of an object of mass $\sim 10 M_{\odot}$.
As for the short GRB 090510, 
the parameter shift appears to be related to the decrease of $r_a/\Gamma_a$,
from $\sim 3 \times 10^7\;$cm to $\sim 3 \times 10^6\;$cm, on a timescale of $t_{\rm delay} \sim 0.1\;$s
(Table~\ref{tab:parameters}). The model fit indicates an even smaller 
$r_a/\Gamma_a \sim 10^5\;$cm for the third time-bin.
This might suggest a smaller size of the progenitor and the central
object than those of long GRBs. 

In such a scenario, the delay timescale of the LAT emission $t_{\rm delay}$ is related to the 
global properties of the jets and the progenitor systems, rather than the microphysical processes 
or the simple kinematic (propagation) effects, and then it might scale with the total duration 
of the MeV emission $t_{\rm dur}$. However, in our specific model, 
while $t_{\rm delay} \sim f_c R_*/c$ scales with
the stellar radius $R_*$ (where $f_c$ is the fraction of the collimation-shocked part at
the jet breakout), $t_{\rm dur}$ may depend on the size of the central region of the star $R_c$ 
from which the accretion rate is very high, i.e., $t_{\rm dur} \sim \sqrt{R_c^3/(G M_c)}$,
where $M_c$ is the mass enclosed within $R_c$ \citep[cf.][]{kumar08b}. 
Then $t_{\rm dur}$ may significantly depend on the angular velocity and density profiles of the star,
which could not simply scale with $t_{\rm delay}$.
Observations of LAT GRBs with redshifts determined show that GRB 080916C, GRB 090902B, and 
GRB 090926A \citep{ackermann090926A} have the durations $t_{\rm dur} \sim 15\;$s, $\sim 8\;$s, and 
$\sim 4\;$s, respectively, while the delay timescales are all $t_{\rm delay} \sim 1\;$s.
In our scenario $t_{\rm delay}$ is expected to be distributed more widely when we obtain a larger 
number of LAT bursts.\footnote{GRB 090217A displays no delay, i.e., $t_{\rm delay} \sim 0\;$s 
\citep[while $t_{\rm dur} (1+z) \sim 33\;$s][]{ackermann090217A}. 
The earliest LAT emission may be just the detections of the 
high-energy tail part of the photospheric emission while the UP emission
onset could correspond to the second group of the LAT detections at $>100\;$MeV, $\sim 7\;$s after
the burst trigger.}

The shift from $\eta > \eta_*$ into $\eta < \eta_*$ makes the change of the photospheric luminosity
from $\approx L$ into $\simeq L(\eta/\eta_*)^{8/3}$, so that it would decrease if $L$ was roughly constant.
The observed MeV emission luminosities appear not to decrease (roughly constant for GRB 080916C and 
GRB 090510 while increase by about a factor of two for GRB 090902B), however, so that $L$ should
increase compensate the shift of the parameter regime of $\eta/\eta_*$. Table~\ref{tab:parameters}
shows that $L$ increases from the first to second time-bins by factors of $\sim 3$ and $\sim 4$ 
for GRB 080916C and GRB 090510, respectively, while for GRB 090902B, the increase factor may be 
estimated to be $\sim 2/(0.52)^{8/3} \sim 10$. This may be another parameter tuning required in our model
for LAT GRBs, but it might be a selection effect: GRB jets with non-increasing $L$ could have dim
UP emission which is not detected by LAT.

For bursts whose LAT emission stays bright from $t_{\rm delay}$ after the trigger, as seen in
the three bursts we have studied in detail, the parameter regime is assumed to stay $\eta < \eta_*$ 
in a large fraction of the total duration in our model. 
This means that the radiation efficiencies in those bursts
are not so extremely large compared with the photospheric emission models in which 
$\eta \ga \eta_*$ is assumed over the total duration.
Indeed we can roughly estimate the radiation efficiencies of the three bursts as
$\epsilon_\gamma \sim \Sigma (L_{\rm ph} + L_{\rm up}) t_{\rm bin} / \Sigma L t_{\rm bin}
\sim 0.4$ for GRB 080916C, $\sim 0.2$ for GRB 090902B, and $\sim 0.3$ for GRB 090510.\footnote{The radiation 
efficiency of the jet consisting of multiple shells could be enhanced
by the cross IC scattering between the shells \citep[e.g.,][]{gruzinov00,li10}. These effects
have not been considered in this paper.} 
Such moderately high efficiencies of prompt emission lead to the external shock emission
as bright as the prompt emission, which is consistent with the observations that all the 
three bursts have bright long-lived high-energy emission. For GRB 090902B, its late-time
low-energy afterglow suggests the total isotropic kinetic energy after the prompt phase $\sim 10^{54}\;$erg
\citep{cenko10}, which is much smaller than estimated from the high-energy afterglow $\sim 10^{55}\;$erg
\citep{kumar10}. This might be attributed to the two-components structure of the jet involving a narrow
and bright spot at the line of sight surrounded by a wider and less energetic region \citep{liu10}.
Many early X-ray afterglows observed by {\it Swift} show very steep decays $(F_{\nu} \sim t^{-3})$,
which are not compatible with the bright high-energy afterglows of LAT GRBs, not decaying so rapidly.
This is another issue to be solved, and a simultaneous observation by {\it Fermi}/LAT and 
{\it Swift}/XRT from the early times will be very helpful. 

The above summary indicates that the photospheric emission models may be viable for the temporal
and spectral properties of the MeV and high-energy emission of 
the {\it Fermi}/LAT GRBs as well as the other ordinary GRBs,
although we need the fine tuning of several parameters (see also the following discussion).
Very recently \citet{peer10} showed detailed simulations of 
the emission processes at a dissipation region out of the photosphere of the jet in a model similar
to ours, including photospheric emission, and concluded that such an emission model can explain 
the spectrum of GRB 090902B, supporting our rough analytical model. \citet{ioka10} explores the 
emission from the photosphere and internal shock of the jet with a very high $\eta \sim 10^4 - 10^6$ 
and argues that the high-energy emission of the LAT GRBs may be explained as synchrotron emission
from the internal shock.

An outstanding problem in the photospheric emission models, including ours is that the low-energy 
spectrum of the photospheric emission, usually assumed to be the Rayleigh-Jeans part of the blackbody 
radiation, $\alpha_{\rm ph} = 1$, is much harder than those of the Band components of typical observed 
GRBs widely distributed around $\alpha \sim -1$ \citep{preece00,ghirlanda02,kaneko06} as well as 
those of LAT GRBs shown in Figure~\ref{fig:spec_080916C}, \ref{fig:spec_090902B}, \ref{fig:spec_090510},
and \ref{fig:spec_090510_3}. 
In this paper we have mainly considered the two-shell system like in Figure~\ref{fig:is_scat}, but
the typical duration for the spectral analysis includes many pulses (Note that our model parameters
and the analysis time-bins are 
$t_v (1+z) \la 0.04\;$s and $T_{\rm bin} > 3\;$s for GRB 080916C,
$t_v (1+z) \sim 0.3\;$s and $T_{\rm bin} \simeq 5\;$s for GRB 090902B, and 
$\tilde{t}_v (1+z) \sim 4 \times 10^{-3}\;$s and $T_{\rm bin} \simeq 0.1\;$s for GRB 090510), 
and thus the superposition
of the photospheric emission from the multiple shells with different $\varepsilon_{\rm ph}$ has the
potential of reproducing the observed low-energy spectrum 
$\alpha \ll 1$.\footnote{The idea of the superposition of the multiple shells emission is similar to a
common interpretation of the flat radio spectrum of typical blazars ($F_\nu \propto \nu^{\delta}$
with $\delta \la 0$). The radio synchrotron emission is thought to be highly self-absorbed
($F_\nu \propto \nu^{5/2}$),
and the emission from the multiple shells can reproduce the observed spectral slope 
\citep[e.g.,][]{konigl81,ghisellini09,marscher09}.}
For example, the photospheric emissions
from multiple shells with different $\eta$ and similar $L$, $r_a/\Gamma_a$, and $\mathcal{R}$ have
a relation $L_{\rm ph} \propto \varepsilon_{\rm ph}$ (not the conventional blackbody relation
$L_{\rm ph} \propto \varepsilon_{\rm ph}^4$; see Equation~\ref{eq:ph_high_baryon}), so that their 
superposition appears as a spectrum 
$\varepsilon F_{\varepsilon} \propto \varepsilon^1$, i.e., $\alpha \sim - 1$.\footnote{Our spectral 
modeling have suggested that a correlation $\eta \propto L^{7/16}$ is required in 
order to reproduce the observed correlation $\varepsilon_{\rm ph} \propto L_{\rm ph}^{1/2}$
in the time-binned spectra of GRB 080916C and GRB 090510 (see Section~\ref{sec:case}), 
which should apply for the maximum value of $\eta$ in some range of times.} 
The parameter $\eta$ should be distributed over a range of a factor $\sim 5$ 
to reproduce the low-energy spectrum extending $\sim 1-2$ decades of energies \citep{ghirlanda07}.
However this is just a rough speculation, and we need more detailed consistency checks
within the model and between the model and observations \citep[see also][]{mizuta10}.
The spectral analysis of GRBs with $T_{\rm bin} \ll 0.1\;$s could reveal the intrinsic hard 
spectrum.
Our speculation is not inconsistent with the recent analysis that the spectrum of as small time-bin
as $T_{\rm bin} \sim 0.5\;{\rm s} \sim t_v (1+z)$ in GRB 090902B becomes close to a blackbody, 
while the spectra of smaller time-bins but $T_{\rm bin} > 5\;{\rm s} \gg t_v (1+z)$ in GRB 080916C
keep non-thermal \citep{binbin10,ryde10}.

The synchrotron (and SSC) emission models also have a problem with the low-energy spectral 
index of the Band component. The electrons are required to be cooled so fast that the low-energy
spectral slope should be $\alpha = -3/2$ in simple types of these models \citep[e.g.,][]{ghisellini00}. 
This is usually softer than the observed slopes, so that the superposition effect does not work well.
There have been many models considering the details of the microphysics in the emission sites, but
there is no consensus for this problem yet \citep[e.g.,][]{ghisellini99,panaitescu00,medvedev00,
derishev01,peer06,asano_terasawa09,daigne10}.

The spectral excesses from the Band component below $\sim 40\;$keV in GRB 090902B could be
one of the important discoveries by {\it Fermi}/LAT, although such an excess was not detected by
other detectors extending to sufficiently low energies such as {\it BeppoSAX} and {\it HETE-2} 
\citep[e.g.,][]{ghirlanda07} \citep[but see][]{ryde06}.
If these are real detections, they may be explained in our model as synchrotron or SSC emission from
internal shocks, constraining the values of 
$\epsilon_d \epsilon_B \sim 0.1$ (see Table~\ref{tab:parameters}).
On the other hand, in order for the synchrotron emission not to be prominent in the low-energy 
range in GRB 090510, we require $\epsilon_d \epsilon_B \la 7 \times 10^{-3}$ and $\la 3 \times 10^{-6}$
for the second and third time-bins, respectively. Such a wide range of constrained parameter 
$\epsilon_B$ is similar to the case in the external shock, $10^{-5} \la \epsilon_B \la 10^{-2}$, 
constrained from the late afterglow observations \citep{panaitescu02}, but it is one of the 
fundamental problems for the shock physics to deduce the microphysical parameters $\epsilon_B$ as well as
$\epsilon_e$ and $p$ from the first principles.

The light curves dominated by the synchrotron or SSC emission should be correlated with those dominated
by the UP emission, since they both are produced by the same internal shocks. Such a correlation 
analysis would be useful to distinguish our photospheric emission model from, e.g., other EIC models
\citep{toma09}. If the high-energy emission is produced by up-scattering of soft X-ray emission from
an external source, such as a cocoon ejected from a progenitor star, off the electrons in the jet, 
the light curve of the seed soft X-rays should not be correlated with that of the high-energy component.
We note that the kinematic arguments in Section~\ref{sec:temporal} indicate that a fraction of 
the photospheric emission should also be temporally correlated with the synchrotron, SSC, and UP 
emission in our model.

The synchrotron component could also be detected in the optical band. For the model parameters
for the second time-bin of GRB 090902B shown in Table~\ref{tab:parameters}, the synchrotron 
self-absorption energy in the observer frame is estimated as 
$\varepsilon_{{\rm syn},a}/(1+z) \sim 7\;$eV and then we have a flux $\sim 320\;$mJy 
at $3 \times 10^{14}\;$Hz, which is a very bright optical source even though it suffers a strong 
self-absorption. 
The bright optical prompt emission of some GRBs, being often much brighter than the extrapolation of the 
Band component \citep[e.g.,][]{briggs99,racusin08}, could be attributed to the synchrotron emission 
from the internal shocks in our model.

The spectra of the three LAT GRBs we considered may have spectral breaks at $\varepsilon_{{\rm up},h}$
and at $\varepsilon_{\rm KN}$ in the high-energy range, while a break due to the $e^{\pm}$ pair creation 
in the emission site is estimated to be typically much above those breaks in our model. Detecting a spectral
break in the high-energy range by {\it Fermi}/LAT and/or by the future Cherenkov telescope
such as CTA\footnote{http://www.cta-observatory.org.} would be very helpful to constrain the 
models.\footnote{The distinct high-energy power-law
component of GRB 090926A has been confirmed to have a spectral break in \citet{ackermann090926A} 
accepted after the submission of our paper.}
Polarimetric observations of the prompt emission also have the potential of distinguishing
the photospheric emission models from the synchrotron emission models \citep{fan09}. In the energy 
range $\varepsilon \la 1\;$MeV, which is the target of some planned missions, the photospheric emission 
which consists of the blackbody component and the multiple-scattered blackbody component may have 
very low polarization, while some of the synchrotron emission models predict detectable degrees of
polarization \citep[see also][for general discussion of GRB polarization]{toma09p}.

The case studies in this paper have shown that the photosphere-internal shock model can produce three
types of the spectral shapes of GRB prompt emission: (i) The photospheric emission with the high-energy tail
of the photon index $\beta_{\rm ph} \la -2.5$ makes the Band spectrum up to $\sim 1-10\;$GeV while a dim 
UP emission has a small contribution or just makes the extension of that Band spectrum at $\ga 1-10\;$GeV
(like the first time-bins of GRB 080916C and GRB 090510); (ii) The UP emission is bright but
the combination of the UP and photospheric components mimics a Band function with the high-energy
index of $\beta > \beta_{\rm ph}$ (like the second to fifth time-bins of GRB 080916C); (iii)
The UP emission is bright and makes a high-energy component distinct from a photospheric emission
(like the second time-bin of GRB 090902B and the second and third time-bins of GRB 090510).
The increasing number of bursts being observed in the LAT field of view 
(including bursts with no detections in the LAT energy range) will determine the number 
ratios of the Band-only spectra and the Band plus distinct high-energy component spectra.
Extensive parametric studies would then be able to clarify whether our model with reasonably large
parameter space is consistent with the high-energy parts of general GRB spectra or 
only consistent with limited numbers of them. A very recent analysis of 52 bright {\it Fermi}/GBM
bursts ($\sim 8\;{\rm keV} - 38\;$MeV) show that the number of the bursts that
can be fitted by a Band function with $\beta > -2.4$ is 14 ($\simeq 27\%$) and with $\beta > -2.3$
is 7 ($\simeq 13\%$) \citep{bissaldi11}. This implies that the fraction of bursts with the 
spectral type (ii) is small, and thus we  might not require significant fine tuning of parameters 
for general GRBs.\footnote{The analysis results of the BATSE data ($\sim 30\;{\rm keV} - 2\;$MeV) show that
many spectra are adequately fitted with a cutoff power-law function as well as a Band function
\citep{kaneko06}, so that they are not suitable for constraining the $\beta$ values and our model.}

We have focused on baryonic jets which evolve roughly adiabatically and have applied a simple
analytical formulation of the radiation from the photosphere and internal shock to deduce 
the physical parameters of the jets for each time-bin. The fine tunings of several parameters
appear to be required for the three bursts we have studied, which might suggest other possibilities, 
e.g., non-adiabatic (i.e., significantly dissipative) jets \citep{rees05}, or magnetically-dominated jets 
\citep[e.g.,][]{thompson94,spruit01,lyutikov06,giannios06,fan09,zhang10}. Even in those other
jet models, however, the photospheric emission, its up-scattering effect in the internal
shocks at large radii,\footnote{Even in the jets in which the magnetic
field energy is dominant at the base of the jet, the field energy can be converted into 
the kinetic energy through the jet expansion, leading to the internal shocks at large radius 
\citep[e.g.,][]{granot11}.}
and the change of the radiation properties due to the interaction of the jet with the 
dense environment could be important for some cases.
Thus, the time-binned spectra and delay timescales of the LAT emission onsets could have
the potential of revealing the radial structure of GRB jets.
The difference in the delay timescales of long and short GRBs may
provide clues for understanding the differences between the jets themselves as well as their
progenitors.

\section*{Acknowledgments}
We thank the {\it Fermi} LAT/GBM group members, B.~B.~Zhang, and T.~Sakamoto for useful discussions. 
We also thank the anonymous referee for comments that significantly improved the paper.
We acknowledge NASA NNX09AT72G, NASA NNX08AL40G, and NSF PHY-0757155 for partial support.
XFW was supported by the National Natural Science Foundation of China
(grants 10633040 and 10921063), National Basic Research Program
of China (973 Program 2009CB824800), and the Special Foundation for the 
Authors of National Excellent Doctorial Dissertations of P. R. China by
Chinese Academy of Sciences. PM is grateful for the hospitality of the Institute for
Advanced Study, Princeton, during part of this project.

\bsp

\label{lastpage}

\end{document}